\documentclass[preprint]{elsarticle}

\usepackage{amsmath}
\usepackage{amssymb}
\usepackage{array}
\usepackage{siunitx}
\usepackage{tikz}
\usepackage{ifthen}
\usepackage{calculator}
\usepackage[linesnumbered,lined,boxed,commentsnumbered]{algorithm2e}
\usepackage{float}
\usepackage[final]{changes}
\usepackage{hyperref}

\journal{Computer Physics Communications}

\DeclareMathOperator{\im}{i}
\newcommand{\del}[2]{\ensuremath{\frac{\partial #1}{\partial#2}}}
\newcommand{\eto}[1]{\ensuremath{\mathrm{e}^{#1}}}
\newcommand{\erwartung}[1]{\ensuremath{\left\langle#1\right\rangle}}
\newcommand{\ordnung}[1]{\ensuremath{\mathcal{O}\left(#1\right)}}
\newcommand{\symin}{symplectic integrator}
\newcommand{\Omin}{Omelyan integrator}
\newcommand{\fourin}{4th order integrator}
\newcommand{\HB}{Hasenbusch}
\newcommand{\cnt}{carbon nanotube}

\newcommand{\graphene}[2]{
	\foreach \x in {0,...,#1}
	{
		\MODULO{\x}{2}{\xmod} 
		\foreach \y in {0,...,#2}
		{
			\tikzset{yshift={\y*1.732cm+Mod(\x,2)*0.866cm}, xshift={\x*1.5cm}}
			\begin{scope}
				\draw[line width=2pt] (0,0)--(1,0);
				\ifthenelse{\xmod=0 \OR \y<#2 \AND \x>0}{\draw[line width=2pt] (0,0)--(-0.5, 0.866);}{}
				\ifthenelse{\xmod>0 \OR \y>0 \AND \x>0}{\draw[line width=2pt] (0,0)--(-0.5, -0.866);}{}
				
				\fill[red] (0,0) circle (4pt);
				
				\fill (1,0) circle (4pt);
				\ifthenelse{\xmod=0 \OR \y<#2 \AND \x>0}{\fill (-0.5, 0.866) circle (4pt);}{}
				\ifthenelse{\y > 0 \AND \x>0}{\fill (-0.5, -0.866) circle (4pt);}{}
			\end{scope}
}}}

\SetAlCapSkip{1ex}

\begin{document}

\begin{frontmatter}

\title{Accelerating Hybrid Monte Carlo simulations of the Hubbard model on the hexagonal lattice}
\author[JSC]{Stefan Krieg}
\ead{s.krieg@fz-juelich.de}
\author[IKP,UBonn]{Thomas Luu}
\ead{t.luu@fz-juelich.de}
\author[UBonn]{Johann Ostmeyer}
\ead{ostmeyer@hiskp.uni-bonn.de}
\author[ICL]{Philippos Papaphilippou}
\ead{pp616@ic.ac.uk}
\author[UBonn]{Carsten Urbach}
\ead{urbach@hiskp.uni-bonn.de}
\address[JSC]{J\"ulich Supercomputing Centre, Forschungszentrum J\"ulich GmbH, J\"ulich, Germany}
\address[IKP]{Institute for Advanced Simulation (IAS-4), Institut f\"ur Kernphysik (IKP-3), and J\"ulich Center for Hadron Physics, Forschungszentrum J\"ulich GmbH, J\"ulich, Germany}
\address[UBonn]{Rheinische Friedrich-Wilhelms-Universit\"at Bonn, D-53012 Bonn, Germany}
\address[ICL]{Dept. of Computing, Dept. of Electrical and Electronic Engineering, Imperial College London, London, United Kingdom}

\begin{abstract}
We present different methods to increase the performance of Hybrid Monte Carlo simulations of the Hubbard model in two-dimensions.  Our simulations concentrate on a hexagonal lattice, though can be easily generalized to other lattices. It is found that best results can be achieved using a flexible GMRES solver for matrix inversions and the second order \Omin\ with \HB\ acceleration on different time scales for molecular dynamics. We demonstrate how an arbitrary number of \HB\ mass terms can be included into this geometry and find that the optimal speed depends weakly on the choice of the number of \HB\ masses and their values. As such, the tuning of these masses is amenable to automization and we present an algorithm for this tuning that is based on the knowledge of the dependence of solver time and forces on the \HB\ masses. We benchmark our algorithms to systems where direct numerical diagonalization is feasible and find excellent agreement.  We also simulate systems with hexagonal lattice dimensions up to $102\times 102$ and $N_t=64$.  We find that the \HB\ algorithm leads to a speed up of more than an order of magnitude.
\end{abstract}


\end{frontmatter}

\section{Introduction} 
The Hubbard model has been used to describe many interesting phenomena relevant to low-dimensional solid-state materials, such as Mott insulating behaviour, anti-ferromagnetic order, etc. For the one-dimensional system, under the Bethe Ansatz~\cite{Lieb:1968zza}, there is no first-order transition between an insulating and conducting state. For two-dimensions (and higher), such a transition is expected to occur at some critical coupling, but to date the exact value of this coupling is unknown.  For weak coupling below the critical coupling, standard perturbative many-body techniques have been applied to this system \cite{Giuliani2009}.  But for coupling near or above the critical coupling, very little is known \emph{analytically} about the Hubbard model in these higher dimensions (see, e.g. \cite{PhysRevB.92.045111}).  What is known has been deduced from numerical simulations, such as via Random Phase Approximation \cite{doi:10.1143/JPSJ.70.1483}, mean field theories \cite{RevModPhys.68.13}, cluster theories \cite{FANG20152230}, or functional renormalization group techniques \cite{PhysRevLett.100.146404,PhysRevLett.100.156401}. Recent calculations on modest system sizes have utilized Quantum Monte Carlo (QMC) techniques\cite{doi:10.1143/JPSJ.70.1483,Sorella2012,Buividovich:2016qrg,Beyl:2017kwp,Lin2015}.  These QMC calculations employ either the Hybrid Monte Carlo (HMC) algorithm \cite{Duane:1987de}, a method used extensively in lattice quantum chromodynamics (QCD)\cite{Armour:2009vj,Drut:2009aj,Smith:2014tha,luu}, or the BSS algorithm\cite{PhysRevD.24.2278} which is more prevalent in the condensed matter community \cite{MengGraphene,Otsuka:2015iba}. In \cite{Otsuka:2015iba} lattices as large as $36\times 36$ were simulated, providing the most precise value of the critical coupling in which a Mott insulating phase occurs.  The largest hexagonal lattice sizes simulated to date are $48\times 48$ with $N_t=80$ timesteps \cite{Stauber:2017fuj}\footnote{The total number of lattice sites for an $n\times m$ hexagonal lattice is $2nm$.  The total dimension of a simulated system includes the number of timesteps $N_t$ and is $2nmN_t$.}.

Many of these studies have been further motivated by the discovery of graphene, first isolated in 2004 by Andre Geim and Konstantin Novoselov. Consisting of carbon ions within an hexagonal lattice, graphene's unique electrical, thermal, and mechanical properties make it a prime candidate (as well as its derivatives, e.g. carbon nanotubes, ribbons, bi-layers) for next generation electrical devices. Numerical simulations of this system (via the Hubbard model and Hamiltonians with more sophisticated electron-electron interactions\cite{Brower:2011av}) will play an essential role in quantifying graphene's (and its allotropes') electrical properties within macroscopic volumes.  The ability to simulate these systems at these scales is, therefore, of prime importance.

The aim of this work is to describe our investigations for speeding up HMC simulations of the Hubbard model on an hexagonal lattice, as well as systems based off this geometry, such as graphene and \cnt s.  By utilising techniques originally developed in the lattice QCD community, we are able to obtain a speedup of more than an order of magnitude in some cases.  This has allowed us to simulate on system sizes as large as $102\times 102$ using single-node runs.  In the following section we give a cursory description of the system and its underlying Hamiltonian. Section~\ref{sect:solver} and~\autoref{sect:integrators} describe our solvers\footnote{We note the recent development of an efficient direct solver via Shur complement~\cite{Ulybyshev:2018dal}. However our work targets system sizes that are much larger than the lattices for which the Schur complement solver is feasible. Ref.~\cite{Ulybyshev:2018dal} argues that a standard CG solver becomes preferable to Shur complement for systems larger than approximately $(70 \times 70)$ unit cells. As our algorithm is much more efficient than standard CG solver we expect it to be faster than Schur complement at even smaller lattices.}
and integrators used when integrating our equations of motions, respectively. In \autoref{sect:multi-scale} we describe our multi-shift solvers, where we provide the scaling and dependence of our algorithms with the \HB\footnote{\HB\ acceleration is an exact preconditioning scheme first introduced in \cite{hasenbusch}.}\ masses.  The following section provides our algorithm for the autotuning of these masses, as well as other relevant parameters to our simulations.  We demonstrate the speedup of our simulations due to the \HB\  solvers in \autoref{sec_results}.  Here we also benchmark our simulation to the exact solutions from a 4-site problem, ensuring that our solver implementations simulate correct physics. We end with a recapitulation of our findings in \autoref{sect:conclusions}.

\section{The system}
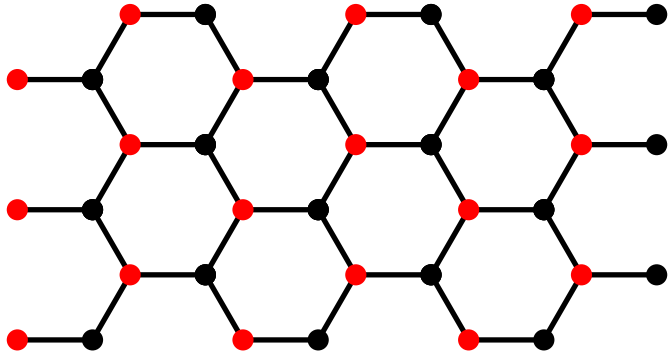
\begin{figure}
	\centering
	\begin{tikzpicture}
	\graphene{5}{2}
	\end{tikzpicture}
	\caption{Honeycomb lattice of graphene. One red and one black adjacent points form a basis of the triangular Bravais lattice.}
	\label{graphene_fig}
\end{figure}
The honeycomb lattice\added{, i.e. a lattice of rhombus-shaped unit cells with a basis of two atoms,} consists of two underlying triangle lattices as depicted in figure~\ref{graphene_fig}. 
This bipartite structure means that any site of one sublattice has three nearest neighbours from the other sublattice. The minimal unit cell, from which the hexagonal lattice is tessellated, consists of two sites A and B, each coming from one of the sublattices. For this work, the orientation of the unit cell is chosen in such a way that the distance between A- and B-site is exactly $\left(a\,,\: 0\right)^T$ where $a$ is the lattice spacing.

In order to investigate the basic properties of the Hubbard model in this geometry, Hybrid Monte Carlo\footnote{Sometimes referred to as Hamiltonian Monte Carlo} (HMC) \cite{Duane:1987de} simulations are used. The underlying Hamiltonian, after Hubbard-Stratonovich transformation and introduction of pseudofermions\cite{luu}, is 
\begin{equation}
\mathcal{H}=\frac{1}{2} \phi^T\frac{1}{\delta U}\phi +\chi^\dagger\left(MM^\dagger\right)^{-1}\!\chi +\frac{1}{2}\pi^T\pi\label{hamiltonian}
\end{equation}
where $\pi$ is the real momentum field, $\phi$ is the real Hubbard field, $\chi$ is a complex pseudofermionic vector field, $\delta$ is the step size in euclidean time dimension weighted by the inverse temperature $\beta$ of the system, $U$ is the Hubbard onsite interaction, and $M$ is the fermion operator (see ref.~\cite{luu}) with
\begin{equation}
\begin{split}
M^{AA}_{(x,t)(y,t')}&=\delta_{xy}\left(-\delta_{tt'}+\left(\eto{\im\phi_{x,t}}-\delta m_s\right)\delta_{t-1,t'}\right)\\
M^{BB}_{(x,t)(y,t')}&=\delta_{xy}\left(\delta_{tt'}-\left(\eto{-\im\phi_{x,t}}-\delta m_s\right)\delta_{t+1,t'}\right)\\
M^{AB}_{(x,t)(y,t')}=M^{BA}_{(x,t)(y,t')}&=-\delta\kappa\delta_{\erwartung{x,y}}\delta_{t,t'}\,.
\end{split}\label{ferm_op}
\end{equation}
Here $\kappa$ is the hopping parameter, $\delta_{\erwartung{x,y}}$ denotes nearest neighbour connections, and $m_s$ is the staggered mass. \added{A nonzero $m_s$ corresponds to a broken symmetry between A- and B-sites, as it introduces a magnetic bias where the spin-up-electrons favour A-sites and spin-down-electrons favour B-sites or the other way around depending on the sign of $m_s$.} All our calculations here are performed with \mbox{$m_s=0\,.$}(In \cite{Ulybyshev:2017hbs} it was pointed out that certain fermion operator discretizations with $m_s=0$ suffer from ergodicity issues.  We stress that our discretization in eq.~\eqref{ferm_op} does not suffer from these same ergodicity issues.) Details related to the derivation of this Hamiltonian can be found in ref.~\cite{luu} and~\cite{hmc_details}, where also an implementation of the standard HMC algorithm to this system was discussed. \added{For an $n\times\ m$ hexagonal lattice, we apply periodic boundary conditions along the oblique axes, as depicted in fig.~4 of \cite{1126-6708-2009-06-060}.}  We note that for graphene and \cnt\ systems, $\kappa\sim \SI{2.7}{\eV}$ and $a=1.42$~\AA \ \cite{Saito1998}.

\added{The Hamiltonian~\eqref{hamiltonian} describes two flavours of spinless fermions, which can be interpreted either as a single electron species with spin one half or as ``particles'' and ``holes''. In our formalism the latter interpretation has been chosen.}

We outline the basic principles of the standard HMC algorithm. The `conjugate momenta' field $\pi$ and auxiliary complex field $\rho$ are sampled according to Gaussian distributions $\eto{-\pi^2/2}$and $\eto{-\rho^\dagger\rho}$, respectively. Then the pseudofermionic field is obtained as $\chi=M\rho$. With these starting parameters and an initial Hubbard field $\phi$ a molecular dynamics (MD) trajectory is calculated and the result is accepted with the probability $\text{min}\!\left(1,\eto{-\Delta \mathcal{H}}\right)$. $\Delta \mathcal{H}$ is the difference in energy resulting from the molecular dynamics. 

\section{Solver\label{sect:solver}}
Note that equation \eqref{hamiltonian} involves a matrix ``inversion'' via the matrix equation $\left(MM^\dagger\right)\eta=\chi$. This equation must be solved for $\eta$ very often, making it beneficial to optimize the linear solver in this case. Observe that $MM^\dagger$ is hermitian and positive definite. This allows us to use the Conjugate Gradient (CG) method (see e.g. ref.~\cite{templates}) to solve the linear system involved. The CG ``solver'', with relative tolerance set to $10^{-8}$ (i.e. squared tolerance $10^{-16}$), and implemented in \texttt{C++} with double precision was used initially to calculate $\eta$.

As a first step to increase the performance another CG solver was added with single precision arithmetic. This single precision CG (spCG) has been used as a preconditioner for the double precision CG. As the spCG with a precision target is a non stationary solver, we have to use the flexibly preconditioned CG (fCG) method\cite{Notay_flexible_CG}. 
We have tested this ``mixed precision'' algorithm on systems of different sizes. On larger systems it led to a speed up of about factor 1.9, thus using nearly the full advantage of factor two between the bandwidths (and peak flop rates) of single and double precision arithmetics. As a next step we replaced the (``outer'') fCG solver by a flexible Generalized Minimal Residual (fGMRES) solver from ref.~\cite{fgmres}. After some fine tuning described in \mbox{\ref{app_fgmres}}, a speed up of factor 2.2 compared to the original solver could be achieved for systems with a total dimension $2L^2N_t$ larger than $\sim\num{100000}$. 

On very large systems one encounters an additional advantage of the spCG preconditioner. The size of the L3-cache is limited and eventually will be too small to store all the vectors required for the solver in double precision. Using single precision arithmetics gives an additional factor of two in the system size until the memory requirements of the spCG preconditioner exceeds the size L3-cache as well. On the compute nodes we used for testing ($2\,\times\,$Intel(R) Xeon(R) CPU E5-2680, $\SI{40}{MB}$ Cache per node), this leads to a speed up of more than factor 4 instead of the expected factor two. 
For more details on the performance see section~\ref{sec_results}.

\section{Integrators\label{sect:integrators}}
In order to improve the calculation of the MD trajectory in the HMC simulations, we tested several different reversible, symplectic integrators\footnote{We confined ourselves to reversible, symplectic integrators to ensure detailed balance in our HMC algorithm\cite{Duane:1987de}. Reversibility is ensured by symmetry which excludes odd order integrators.}. An integrator evolves a system of coordinates $\phi(t)$ and momenta $\pi(t)$ depending on the force $F(\phi)$ by a time step $\Delta t$. Any integrator which is linear in $\Delta t$ follows the rules:
\begin{align}
\phi_i&=\phi_{i-1}+c_i\,\Delta t \,\pi_{i-1}\\
\pi_i&=\pi_{i-1}-d_i\,\Delta t \,F(\phi_i)
\end{align}
Here $(\phi_0,\pi_0)=(\phi(0),\pi(0))$ and $(\phi_n,\pi_n)=(\phi(\Delta t),\pi(\Delta t))$ with $i\in\{0,\dots,n\}$. In addition the condition $\sum_i c_i=\sum_i d_i =1$ has to hold.

The most simple symmetric, thus reversible, symplectic integrator is the ``leapfrog'' integrator. The leapfrog is a second order method ($c_1=c_2=\frac{1}{2}$, $d_1=1$, $d_2=0$ in the velocity version).

\subsection{Second order Omelyan integrator}
In ref.~\cite{omelyan} it was shown that the second order integrator using $c_1=c_3=\zeta$, $c_2=1-2\zeta$, $d_1=d_2=\frac{1}{2}$, $d_3=0$ with $\zeta\approx \num{0.193}$, even though it requires two force calculations, is more effective as it minimizes the leading order error term significantly allowing for a coarser time step. The error can be estimated by expanding the evolution of two non commutative operators $a$ and $b$\added{ (in our case they correspond to the evolution of $\phi$ and $\pi$ respectively)} in the form
\begin{equation}
\eto{(a+b)h}=\eto{a\zeta h}\eto{bh/2}\eto{a(1-2\zeta)h}\eto{bh/2}\eto{a\zeta h}+Ch^3+\ordnung{h^4}
\end{equation}
and minimizing $C$. \added{In general, when no assumptions are made about the operators $a$ and $b$, then} the commutator $[a,b]$ is not known\added{.  In this case } Omelyan used an Euclidean norm spanned by the commutators $[a,[b,a]]$ and $[b,[b,a]]$:
\begin{align}
C&=\added{c_1} [a,[b,a]] + \added{c_2} [b,[b,a]]\\
\Rightarrow ||C||&=\sqrt{\added{c_1}^2+\added{c_2}^2} .
\end{align}
The minimum has the numerical value $||C||\approx 0.009$ \added{when $\zeta=0.193$ (when $\zeta=0$ or $1/2$ one obtains the original leapfrog integrator with $||C_0||\approx 0.09$).} As the \Omin\ uses two force calculations per step the error has to be compared at double step size leading to the error ratio
\begin{equation}
\frac{\epsilon_\text{Omelyan}}{\epsilon_\text{leapfrog}}\approx\frac{0.009}{0.09}2^2=0.4
\end{equation}
where $\epsilon$ denotes the different errors.

\added{We tested different values of $\zeta$ in our simulations with the \Omin\  integrator and found the theoretically derived value $\zeta=0.193$ to be optimal, resulting in a factor of two to three speed-up compared to the generic method.}

\subsection{Higher order integrators}
A detailed list of symmetric \symin s can be found in ref.~\cite{omelyan2003}. Several \fourin s were analysed in our molecular dynamics calculations. As our goal was not to obtain extremely precise solutions\footnote{The accept/reject step renders the algorithm exact even for finite precision integrators\cite{Duane:1987de}.} but rather to achieve a good acceptance rate with minimal cost, we did not investigate yet higher order integrators. The acceptance is weighted by the Boltzmann factor $\eto{-\Delta \mathcal{H}}$ where $\Delta\mathcal{H}$ is the dimensionless deviation in energy accumulated over one trajectory. Targeting an acceptance of about \num{0.66} translates to $\Delta\mathcal{H}\approx \num{0.42}$.

We found the force gradient integrator as described in ref.~\cite{force_grad} most promising. The algorithm reads
\begin{equation}
\begin{split}
\pi_1&=\pi_0-\frac{1}{6}\Delta tF(\phi_0)\\
\phi_1&=\phi_0+\frac{1}{2}\Delta t\pi_1\\
\varphi&=\phi_1-\frac{1}{24}\Delta t^2F(\phi_1)\\
\pi_2&=\pi_1-\frac{2}{3}\Delta tF(\varphi)\\
\phi_3=\phi_2&=\phi_1+\frac{1}{2}\Delta t\pi_2\\
\pi_3&=\pi_2-\frac{1}{6}\Delta tF(\phi_2)
\end{split}\label{force_cal}
\end{equation}
and has an error scaling in $\ordnung{\Delta t^4}$ with three force calculations per iteration.
\begin{figure}[h!]
	\centering
	\includegraphics[angle=270, width=0.9\textwidth]{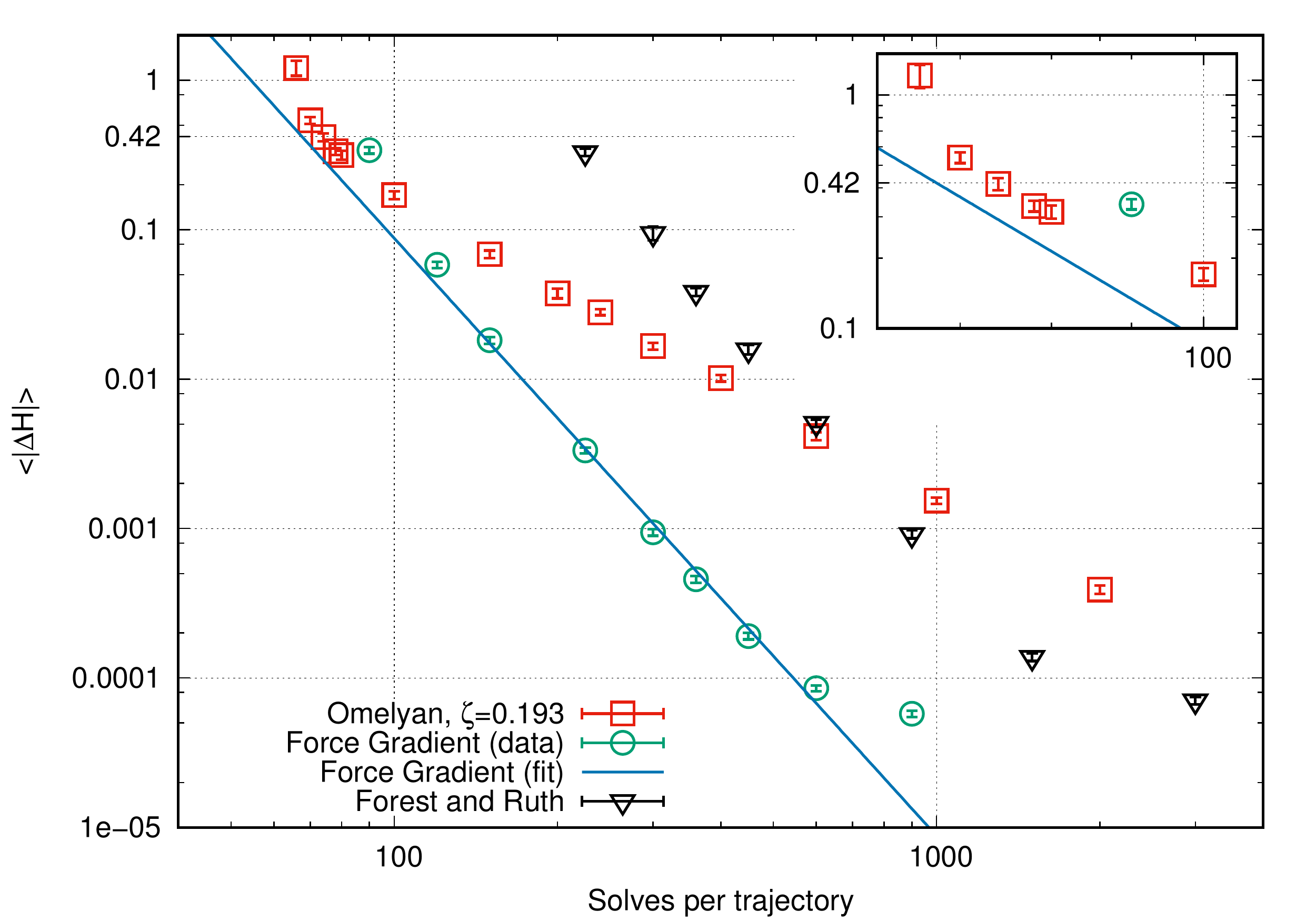}
	\caption{Average absolute deviation of the energy after one trajectory $\left\langle\left|\Delta \mathcal{H}\right|\right\rangle$ for different integrators depending on the cost. The Omelyan integrator requires two force calculations per step, both \fourin s require three. The blue line shows a fit of the error of the force gradient integrator.}
	\label{error_scaling_comp_32}
\end{figure}
Figure~\ref{error_scaling_comp_32} shows the average absolute deviation $\left\langle\left|\Delta \mathcal{H}\right|\right\rangle$ made by different integrators for an arbitrarily chosen system. In the region of interest with the error approximately $\num {0.42}$ the force gradient integrator is comparable to the \Omin. However, the problem here is the well known divergence of the error with slight increase of $\Delta t$ for both algorithms. The fit alone thus does not tell which of the integrators is preferable. In this case the \Omin\ seems to be slightly more stable in the considered region but the advantage gained by using one or the other integrator is negligible.

Our findings suggest that it really does not matter whether one uses the \Omin\ or the force gradient integrator. From all inspected integrators these two are the best (compare exemplary the ``standard'' \fourin\ by Forest and Ruth~\cite{forest_ruth} in figure~\ref{error_scaling_comp_32}) and we decided to use the \Omin\ because it is more flexible for modifications as described in the following sections.
In addition one should mention that due to the improvements stated in this article the calculation of the molecular dynamics is not the bottleneck any more. We spend at least the same amount of time on the measurements after having calculated the trajectories.

\section{Multi-scale integration\label{sect:multi-scale}}
So far, we have discussed improvements to the linear solvers and to the integrators used in the computation of the MD trajectory. The integration follows Hamilton's equations of motion
\begin{eqnarray}
\dot{\phi}^T=&\del{\mathcal{H}}{\pi}&=\pi^T\\
\dot{\pi}^T=&-\del{\mathcal{H}}{\phi}&=-\phi^T\frac{1}{\delta U}+2\,\mathrm{Re}\!\left(\eta^\dagger\del{M}{\phi}\xi\right)\label{p_dot}
\end{eqnarray}
with the two complex vector fields
\begin{equation}
\eta=\left(MM^\dagger\right)^{-1}\chi\,,\;\;\xi=M^\dagger\eta\,.\label{eta_xi}
\end{equation}
The computational costs are dominated by the linear solver. Further improvements to the algorithm are possible by realizing that the rate of convergence of the linear solver and thus the speed of the MD integration is given by the condition number of the fermion matrix, i.e. it is dominated by its low eigenmodes. The algorithmic performance of the solver has already been improved and, while it is believed that there is still room for optimization, here a second approach is investigated.

\subsection{Splitting off the small eigenvalues of the fermion matrix}
In ref.~\cite{hasenbusch} Hasenbusch proposed to replace the pseudofermionic term in the Hamiltonian (eqn.~\ref{hamiltonian}) by two terms:
\begin{equation}
\chi^\dagger\left(MM^\dagger\right)^{-1}\!\chi\mapsto\chi_1^\dagger\left(MM^\dagger+\mu^2\right)^{-1}\!\chi_1+\chi_2^\dagger\left(\frac{MM^\dagger}{MM^\dagger+\mu^2}\right)^{-1}\!\chi_2
\end{equation}
This is possible because the determinant
\begin{align}
\det\left(MM^\dagger\right)&=\int\mathcal{D}\chi\mathcal{D}\chi^\dagger\, \eto{-\chi^\dagger\left(MM^\dagger\right)^{-1}\!\chi}\\
&=\int\mathcal{D}\chi_1\mathcal{D}\chi_1^\dagger\mathcal{D}\chi_2\mathcal{D}\chi_2^\dagger\,\eto{-\chi_1^\dagger\left(MM^\dagger+\mu^2\right)^{-1}\!\chi_1-\chi_2^\dagger\left(\frac{MM^\dagger}{MM^\dagger+\mu^2}\right)^{-1}\!\chi_2}
\end{align}
does not change under the given transformation.
In the case of lattice QCD, this procedure replaces the original ill-conditioned matrix with one that has its low modes shifted away from zero by the factor $\mu^2$ without changing the high modes significantly, while correcting for the change with an additional term in the Hamiltonian. The relation is exact and if, as it turns out to be, the second term is sub-dominant, i.e. leads to a smaller MD force than the first one, it can be included in the MD integration using a larger time-step, hence speeding up the overall simulation. 

With, e.g. Wilson type fermions, adding the shift $\mu^2$ to the matrix $MM^\dagger$ in the determinant is easily achieved by adding a ``twisted'' mass $\mu$ to the fermion operator $M$. In our case this approach does not work because our fermion matrix lacks any ``spinor'' structure.
However, we are free to add a non-zero staggered mass $m_s$, as shown in eq.~\eqref{ferm_op}. This leads to
\begin{equation}
\chi^\dagger\left(MM^\dagger\right)^{-1}\!\chi\mapsto\chi_1^\dagger\left(M_\mu M_\mu^\dagger\right)^{-1}\!\chi_1+\chi_0^\dagger\left(\frac{MM^\dagger}{M_\mu M_\mu^\dagger}\right)^{-1}\!\chi_0
\end{equation}
with the notation
\begin{equation}
M=M(m_s=0)\,,\;\;M_\mu=M(m_s=\mu)\,.
\end{equation}
The generalization to $n$ masses $\mu_0,\dots,\mu_{n-1}$ reads
\begin{align}
&\chi^\dagger\left(MM^\dagger\right)^{-1}\!\chi\mapsto\\
&\chi_{n}^\dagger\left(M_{\mu_{n-1}} M_{\mu_{n-1}}^\dagger\right)^{-1}\!\chi_{n}+\sum_{i=1}^{n-1}\chi_i^\dagger\left(\frac{M_{\mu_{i-1}}M_{\mu_{i-1}}^\dagger}{M_{\mu_{i}} M_{\mu_{i}}^\dagger}\right)^{-1}\!\chi_i+\chi_0^\dagger\left(\frac{MM^\dagger}{M_{\mu_0} M_{\mu_0}^\dagger}\right)^{-1}\!\chi_0\,.\nonumber
\end{align}
The $\chi_n$-term can now be treated exactly as the standard pseudofermionic $\chi$-term in the original algorithm by replacing $M$ by $M_{\mu_{n-1}}$. In contrast the $\chi_i$-terms with $i<n$ deserve additional treatment.\\
First of all the creation of $\chi_i$ now involves an inversion. After generating $\rho_i$ from $\eto{-\rho_i^\dagger\rho_i}$ one obtains $\chi_i$ from
\begin{align}
&\chi_i=M_{\mu_i}^{-1}M_{\mu_{i-1}}\rho&=&M_{\mu_i}^\dagger\left(M_{\mu_i} M_{\mu_i}^\dagger\right)^{-1}M_{\mu_{i-1}}\rho\;\;\text{for}\;0<i<n\,,\\
&\chi_0=M_{\mu_0}^{-1}M\rho&=&M_{\mu_0}^\dagger\left(M_{\mu_0} M_{\mu_0}^\dagger\right)^{-1}M\rho\,.
\end{align}
In addition one has to consider the changes in eq.~\eqref{p_dot} where using $\del{M_{\mu_i}}{\phi}=\del{M}{\phi}$ leads to
\begin{eqnarray}
\dot{\pi}^T&=&F_\phi+\sum_{i=0}^{n}F_{\chi_i}\\
F_\phi&=&-\delta\phi^TV^{-1}\\
F_{\chi_n}&=&2\,\mathrm{Re}\!\left(\eta_n^\dagger\del{M}{\phi}\xi_n\right)\\
F_{\chi_i}&=&2\,\mathrm{Re}\!\left(\eta_i^\dagger\del{M}{\phi}\left(\xi_i-\chi_i\right)\right)\;\;\text{for }i<n\label{f_chi_i}
\end{eqnarray}
with $\eta_n$ and $\xi_n$ obtained like in eq.~\eqref{eta_xi} with $M_{\mu_{n-1}}$ instead of $M$. The terms in eq.~\ref{f_chi_i} are defined as follows:
\begin{eqnarray}
\eta_i&=&\left(M_{\mu_{i-1}}M_{\mu_{i-1}}^\dagger\right)^{-1}M_{\mu_{i}} \chi_i\\
\xi_i&=&M_{\mu_{i-1}}^\dagger\eta_i = M_{\mu_{i-1}}^{-1}M_{\mu_{i}} \chi_i\label{def_xi_i}
\end{eqnarray}
Here $M_{\mu_{-1}}\equiv M$ has been used for simplicity. Note that in the case of $\mu_{i}=\mu_{i-1}$ this leads to $\xi_i=\chi_i$ and thus $F_{\chi_i}=0$, leading to the reduced set of equations with one less pseudofermion field. It can be assumed that $F_{\chi_i}$ changes continuously in $\mu_i$ and $\mu_{i-1}$ (see below for more details). Therefore, this part of the force will be weak for small $\left|\mu_{i}-\mu_{i-1}\right|$. On the other hand $M_{\mu_i}$ is better conditioned than $M$ because the eigenvalues are bound from below by $\mu_i$. This leads to a splitting of the original force into $n+1$ terms of which some are easy to calculate and the other ones can be chosen small.

\subsection{Different time scales}
One can use this result in the same way as proposed in ref.~\cite{urbach}. The force $F_{\chi_0}$ can be treated with less accuracy by putting it on a different time scale. One can do this for each successive $F_{\chi_1}$, $F_{\chi_2}$, and so on. Let us for simplicity first consider the case with only one staggered mass term. We use the \Omin\ with the parameter $\zeta=0.193$ for the calculation of the trajectory. Let $T(t)$ denote the evolution of the system by one trajectory of length $t$, $T_\pi(\Delta t)$ denote the evolution of $\phi$ over the time $\Delta t$, and $T_{\phi,\chi_0,\chi_1}(\Delta t)$ denote the evolution of $\pi$ corresponding to the forces $F_{\phi,\chi_0,\chi_1}$ respectively. Then the na\"ive \Omin\ reads
\begin{multline}
T(\Delta t)=T_{\phi+\chi_0+\chi_1}(\zeta\Delta t)\cdot T_\pi\left(\Delta t/2\right)\\
\cdot T_{\phi+\chi_0+\chi_1}((1-2\zeta)\Delta t)\cdot T_\pi\left(\Delta t/2\right)\cdot T_{\phi+\chi_0+\chi_1}(\zeta\Delta t)
\end{multline}
and $T(t)=T(\Delta t)^{N_\text{MD}}$ where $N_\text{MD}=\frac{t}{\Delta t}$ is the number of time steps in the molecular dynamics. As we want to weight $T_{\chi_0}$ less we can choose a number $N_0$ that divides $N_\text{MD}$ and write
\begin{eqnarray}
T(N_0\Delta t)&=&T_{\chi_0}(N_0\zeta\Delta t)\cdot T_0(\Delta t)^{N_0/2}\nonumber\\
&&\cdot T_{\chi_0}(N_0(1-2\zeta)\Delta t)\cdot T_0(\Delta t)^{N_0/2}\cdot T_{\chi_0}(N_0\zeta\Delta t)\\
T_0(\Delta t)&=&T_{\phi+\chi_1}(\zeta\Delta t)\cdot T_\pi\left(\Delta t/2\right)\nonumber\\
&&\cdot T_{\phi+\chi_1}((1-2\zeta)\Delta t)\cdot T_\pi\left(\Delta t/2\right)\cdot T_{\phi+\chi_1}(\zeta\Delta t)
\end{eqnarray}
where in case of $N_0$ odd we define 
\begin{equation}
T_0(\Delta t)^{1/2}=T_{\phi+\chi_1}(\zeta\Delta t)\cdot T_\pi\left(\Delta t/2\right)\cdot T_{\phi+\chi_1}((1-2\zeta)\Delta t/2)
\end{equation}
if the term stands before $T_{\chi_0}(N_0(1-2\zeta)\Delta t)$ and
\begin{equation}
T_0(\Delta t)^{1/2}=T_{\phi+\chi_1}((1-2\zeta)\Delta t/2)\cdot T_\pi\left(\Delta t/2\right)\cdot T_{\phi+\chi_1}(\zeta\Delta t)
\end{equation}
if it comes later. This manipulation maintains the basic properties of the integrator: It remains symplectic and symmetric.\\
Now $\mu_0$ has to be chosen in such a way that the calculation of $F_{\chi_1}$ is much cheaper than the calculation of $F_{\chi_0}$ but still the magnitude of $F_{\chi_0}$ is much less than of $F_{\chi_1}$. In this case $N_0$ should be chosen as high as possible without significantly reducing the accuracy.\\
For an arbitrary number of \HB\ masses and time scales this scheme can be defined recursively in the same way as shown above. $T_{\phi+\chi_n}$ is the innermost part. It is all together repeated $N_\text{MD}$ times per trajectory, this being $N_{n-1}$ times as often as $T_{\chi_{n-1}}$, $N_{n-2}$ times as often as $T_{\chi_{n-2}}$, and so on, up to $N_0$ times as often as $T_{\chi_0}$. Here obviously $N_{n-1}$ has to divide $N_{n-2}$, $N_{n-2}$ has to divide $N_{n-3}$, $\dots$, $N_0$ has to divide $N_\text{MD}$. Of course it is possible to choose $N_i=N_{i+1}$.

\section{Scaling behaviour of the combined method}
Using the methods described in the previous sections, we have performed simulations for different systems, one of which is shown here in greater detail: a $(15,15)$-chirality nanotube with 15 unit lengths and periodic boundary conditions applied at the ends, divided into $N_t=256$ time slices. \added{Such boundary conditions are defined using an orthogonal basis as depicted in fig.~3 of \cite{1126-6708-2009-06-060}.}\added{ The coupling is $U/\kappa=\num{3.5}$ and the inverse temperature $\beta\kappa=8$.}

\subsection{Dependence of the solver time on $m_s$}
\begin{figure}[h!]
	\centering
	\includegraphics[angle=270, width=0.9\textwidth]{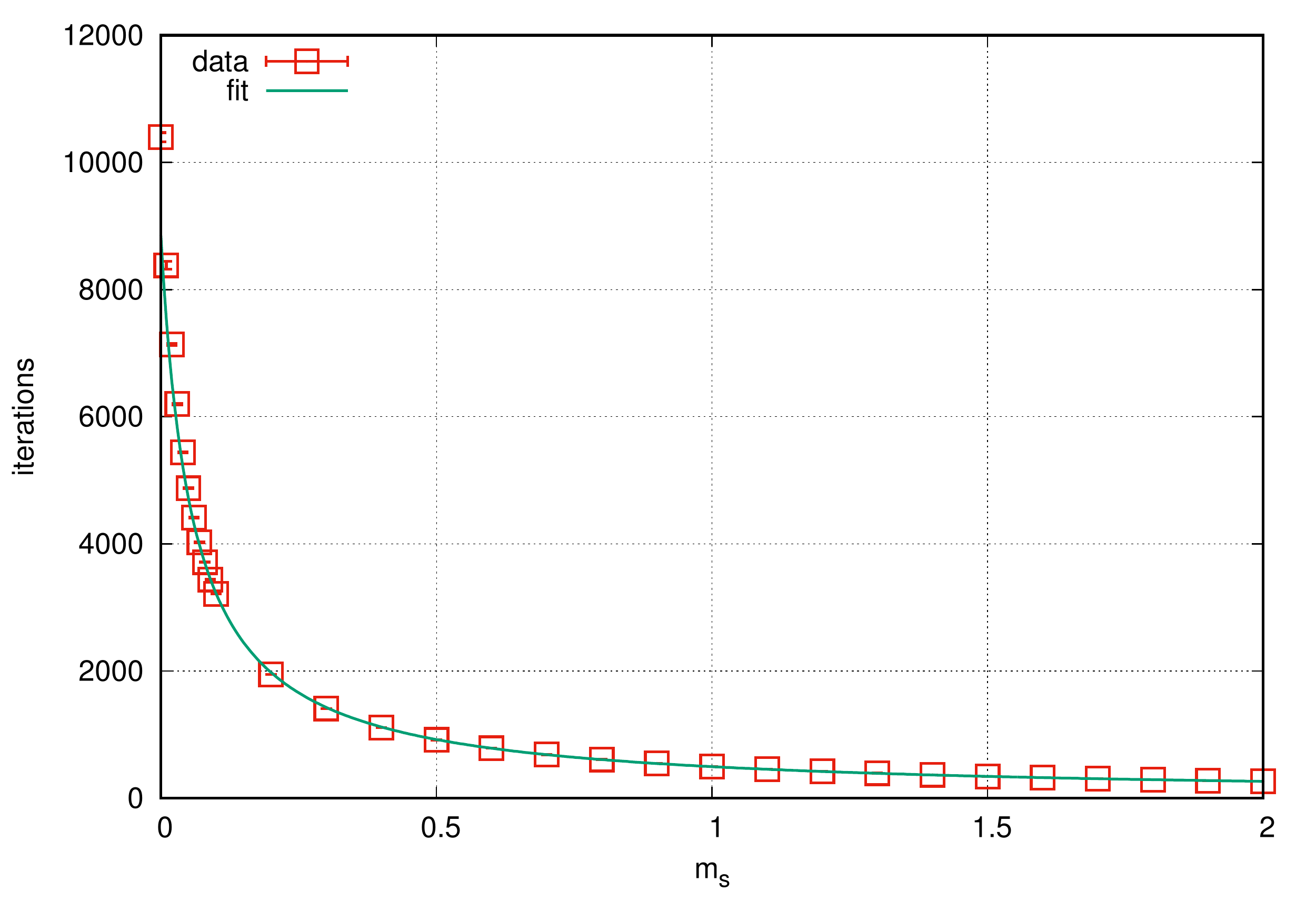}
	\caption{Runtime of one inversion of $M_{m_s}M_{m_s}^\dagger$ depending on the staggered mass $m_s$.}
	\label{time_scaling}
\end{figure}

First of all it is important to understand the scaling of the solver time depending on the staggered mass parameter $m_s$. It is known (e.g. from ref.~\cite{saad_gmres}) that for the flexible Generalized Minimal Residual (GMRES) solver the time scales as
\begin{equation}
t=t_0\frac{\lambda_\text{max}}{\lambda_\text{min}}
\end{equation}
where $\lambda_\text{max}$ is the largest and $\lambda_\text{min}$ the smallest eigenvalue of the matrix that has to be inverted. As $m_s$ shifts the complete eigenspectrum of the matrix this becomes
\begin{equation}
t\left(m_s\right)=t_0\frac{\lambda_\text{max}+m_s}{\lambda_\text{min}+m_s}\,.\label{solver_time_eqn}
\end{equation}
It turns out that \added{even for our nested solvers} this is a good approximation for $\lambda_\text{min}<m_s<\lambda_\text{max}$ as can be seen in figure~\ref{time_scaling}. The depicted fit corresponds to $\lambda_\text{max}=\num{4.1\pm 0.2}$ and $\lambda_\text{min}=\num{0.054\pm 0.002}$.
This functional form describes the data sufficiently well. It has the advantage that it directly gives the highest and the lowest value of the spectrum of the original matrix $MM^\dagger$. These values can directly be used to find a good set of \HB\ masses.

\added{The largest eigenvalue $\lambda_\text{max}$ can be approximated with the non-interacting limit for special cases (e.g. for graphene without chemical potential using the Hubbard model the largest eigenvalue known from the non-interacting limit shows very small deviations from the values with interaction). However it is not known in general and the purpose of the auto--tuner is much more general. We find that it works extremely well for graphene and carbon nanotubes without having to provide any additional information.  We expect it to be applicable even when we add e.g. non--onsite interactions.}

\subsection{Dependence of $F_{\chi_i}$ on $\mu_{i}$ and $\mu_{i-1}$}
\begin{figure}[h!]
	\centering
	\includegraphics[angle=270, width=0.9\textwidth]{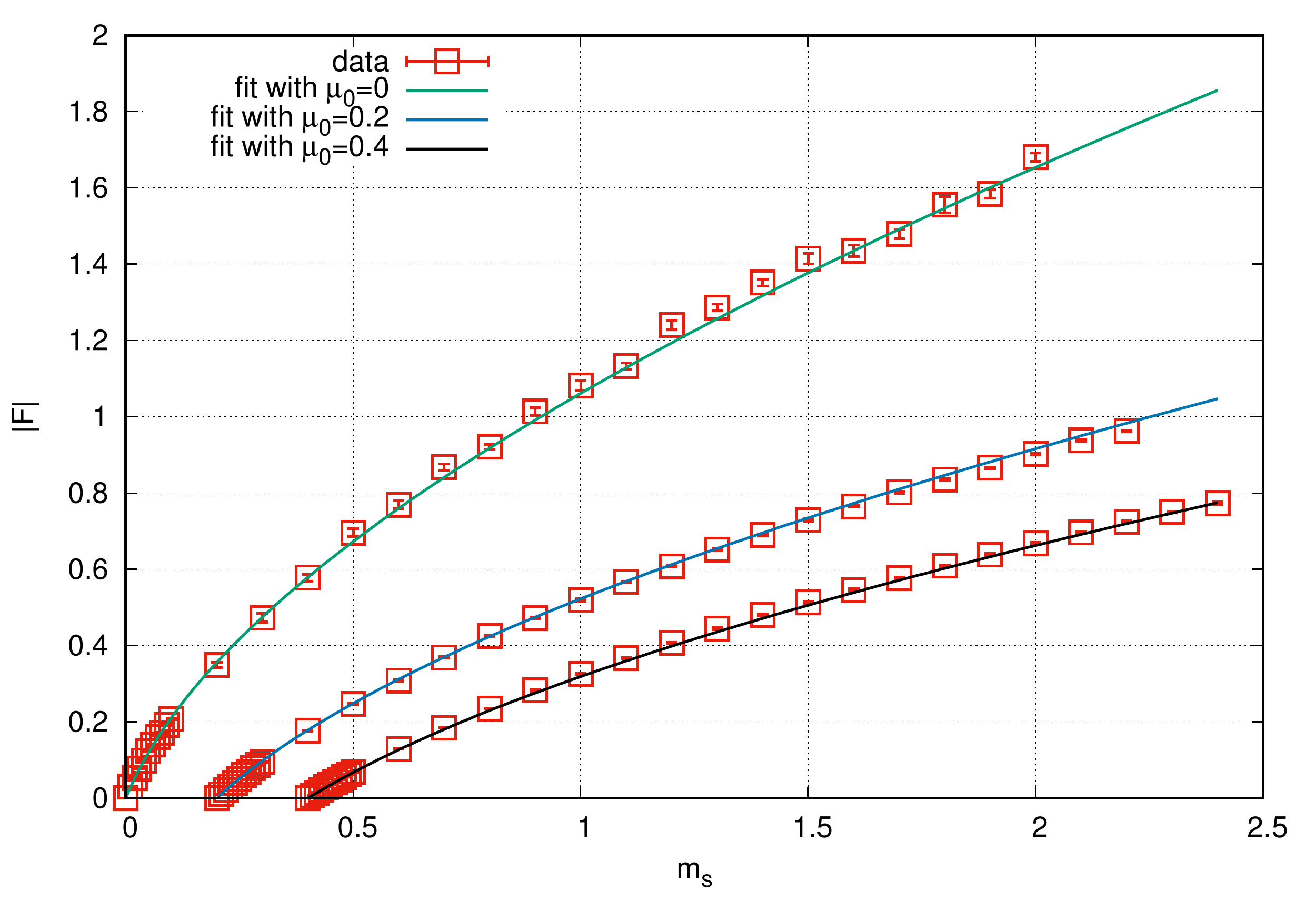}
	\caption{Force $\left|F_{\chi_1}\right|$ depending on the staggered mass $m_s=\mu_1$ for several fixed values of $\mu_0$. We find the fit parameters $a=\num{0.413\pm 0.013}$, $b=\num{0.078\pm 0.026}$ and $c=\num{0.380\pm 0.038}$.}
	\label{small_force_scaling}
\end{figure}

Let us, for simplicity, assume that we simulate with two staggered masses $\mu_0$ and $\mu_1$. As any force term $F_{\chi_i}$ depends only on at most two masses, this consideration is general enough to understand all force terms except for $F_{\phi+\chi_n}$.\\
From eqs.~\eqref{f_chi_i} and~\eqref{def_xi_i} follows
\begin{align}
F_{\chi_1}&\propto\left(M_{\mu_0}^{-1}M_{\mu_1}-1\right)\chi_1\\
&=\left(M_{\mu_0}^{-1}M_{\mu_0+\left(\mu_1-\mu_0\right)}-1\right)\chi_1\label{exact_prop_f_chi_1}\\
&\approx \left(M_{\mu_0}^{-1}\left(M_{\mu_0}+\left(\mu_1-\mu_0\right)\right)-1\right)\chi_1\label{approx_prop_f_chi_1}\\
&=M_{\mu_0}^{-1}\left(\mu_1-\mu_0\right)\chi_1
\end{align}
where, between eqs.~\eqref{exact_prop_f_chi_1} and~\eqref{approx_prop_f_chi_1}, we assumed that the staggered mass term can be placed on the time diagonal. This approximation is valid as it produces the correct continuum limit. Thus for small differences $\mu_1-\mu_0$ the norm of the force should be proportional to the absolute value of this difference:
\begin{equation}
\left|F_{\chi_1}\right|\propto\left|\mu_1-\mu_0\right|
\end{equation}
Higher order influences coming from the matrix products have not been considered here. To take them into account one first has to pay attention to the fact that the force should be symmetric under exchange of $\mu_0$ and $\mu_1$. Empirically we have found that
\begin{equation}
\left|F_{\chi_1}\right|=a\frac{\left|\mu_1-\mu_{0}\right|}{\left(\left(b+\mu_0\right)\left(b+\mu_{1}\right)\right)^c}\label{small_force_eqn}
\end{equation}
is a symmetric form obeying the proportionality for small differences and fitting the data sufficiently well (see fig.~\ref{small_force_scaling}).
It has been found that, independent of the system size and geometry, one always has $c\approx 0.4$.
The parameter $b$ is in the same order of magnitude as $\lambda_\text{min}$ found in the previous section. Thus it can be assumed that the factors $b+\mu_0$ and $b+\mu_1$ in the denominator correspond to the minimal eigenvalues of the matrix at staggered masses $\mu_0$ and $\mu_1$ respectively.  

\subsection{Dependence of $F_{\phi+\chi_n}$ on $\mu_{n-1}$}
As the remaining large force term depends only on the last mass $\mu_{n-1}$ we can again consider for simplicity only one staggered mass term $m_s$. Then one can make the ansatz
\begin{equation}\label{eqn:F}
\left|F_{\phi+\chi_n}\right|\approx F_0-\left|F_{\chi_{n-1}}\right|
\end{equation}
for the functional form with some constant $F_0$. However the numerical values of the coefficients have to be fitted independently of the results in the previous section:
\begin{equation}
\left|F_{\phi+\chi_n}\right|=F_0-\added{\tilde{a}}\frac{m_s}{\left(\added{\tilde{b}}\left(\added{\tilde{b}}+m_s\right)\right)^{\added{\tilde{c}}}}\label{large_force_eqn}
\end{equation}
\begin{figure}[h!]
	\centering
	\includegraphics[angle=270, width=0.9\textwidth]{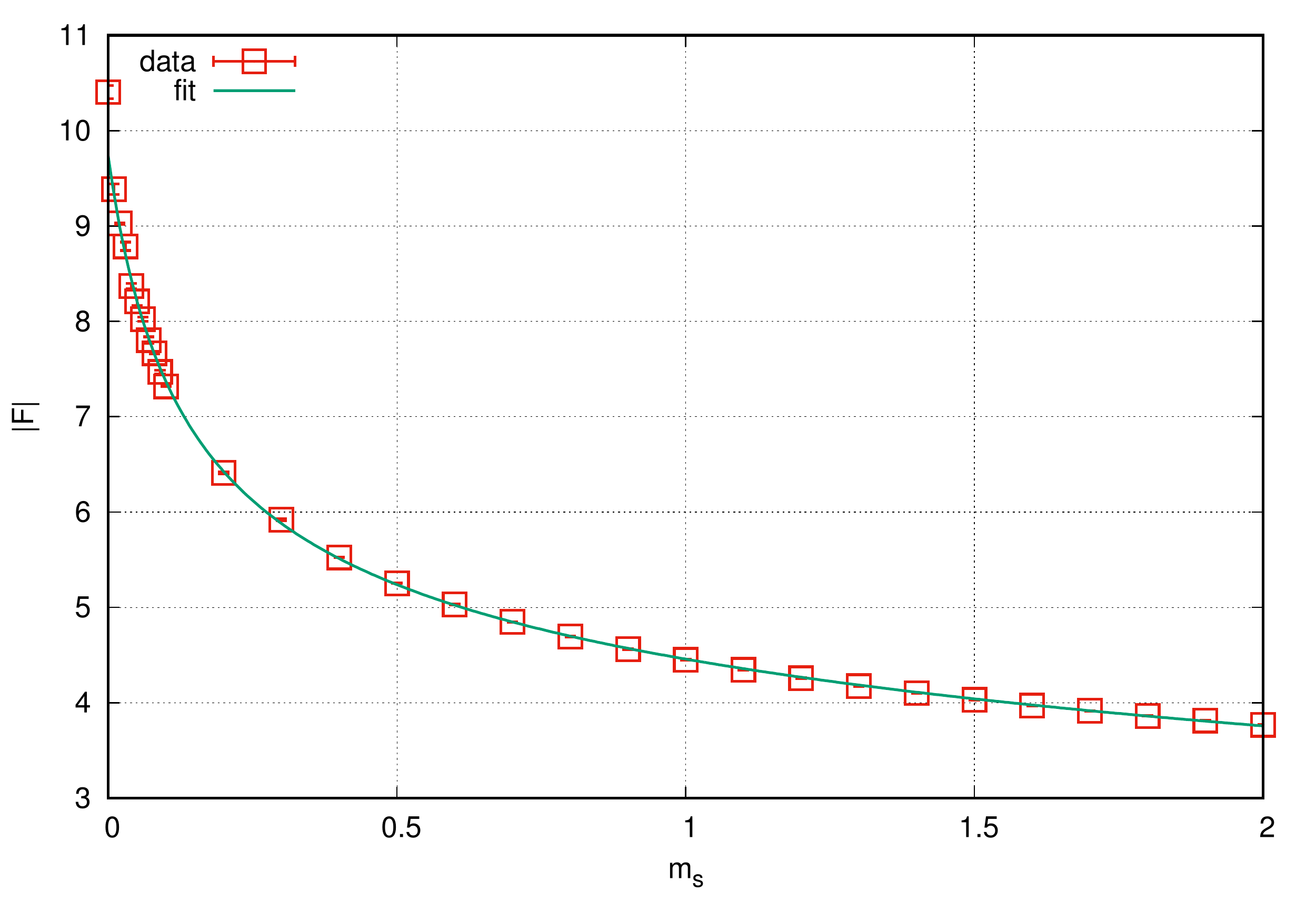}
	\caption{Force $\left|F_{\phi+\chi_n}\right|$ depending on the staggered mass $m_s$ and fitted to \eqref{eqn:F}.}
	\label{large_force_scaling}
\end{figure}
Figure~\ref{large_force_scaling} shows that again for $m_s>\lambda_\text{min}$ the scaling of the force is described very well by the given functional form. One also sees as expected that the values are about an order of magnitude larger than the ones of $\left|F_{\chi_i}\right|$. For the parameters one gets again $\added{\tilde{b}}$ (here $\added{\tilde{b}}=\num{0.0992\pm 0.0056}$) in the same order of magnitude as $\lambda_\text{min}$. $\added{\tilde{c}}=\num{0.789\pm 0.003}$ in this case.

\section{Automatic tuning of the simulation parameters}
In order to explore the whole parameter space, we need to run a large number of independent simulations, at least one for each combination of coupling $U$, temporal lattice spacing $\delta$, number of time steps $N_t$, and system size $L$. Such a large set of independent simulations is hard to tune by hand, and therefore the ability to auto-tune these parameters is essential.  

\subsection{Tuning the HMC step size}
Our auto--tuner executes a series of successive simulations, each time continuing the simulation for a single MD trajectory, until some termination conditions (which the user supplies) are met.\added{ The length of the trajectory is fixed and has to be chosen by the user. It has been set to $t=1$ here for reasons of simplicity yielding $\Delta t=\frac{1}{N_\text{MD}}$.} There is one main loop consisting of two phases. In the first phase, the auto--tuner runs the simulator multiple times with the same parameters until the mean acceptance rate has a small confidence interval, making sure that the acceptance rate observation for the current $N_\text{MD}$ is sufficiently accurate. In the second phase, the auto--tuner collects the observations from all previous iterations and tries to create a new model for the relationship between acceptance rate and $N_\text{MD}$. After the model is created, the $N_\text{MD}$ for the next iteration is selected based on the model’s prediction for yielding a preset acceptance rate, e.g. $p_\text{acc}=0.66$. The loop stops when the parameters have converged or the number of trajectories has reached a specified maximum limit.

\paragraph {First phase -- minimizing the confidence interval}
In the first phase, the auto--tuner tries to get an accurate measurement of the acceptance rate for the current $N_\text{MD}$. In order to do this, a number of samples of trajectories is simulated. Initially, a minimum number of trajectories is required. Next, the confidence interval is updated after each additional trajectory and the phase stops after this measurement gets smaller than a specified limit. The confidence interval is based on the variance of a normally distributed population and then the bounds are minimized if a portion of it lies outside the range between 0 and 1. It is important to note that here we make the assumption that the acceptance rates for the samples of trajectories are normally distributed across the samples and, therefore, it may be subject for further refinement.\added{ In our experience the number of trajectories needed in this phase ranges from five to 30.}

\begin{figure}[H] 
	\centering
	\includegraphics[angle=0, width=0.8\textwidth]{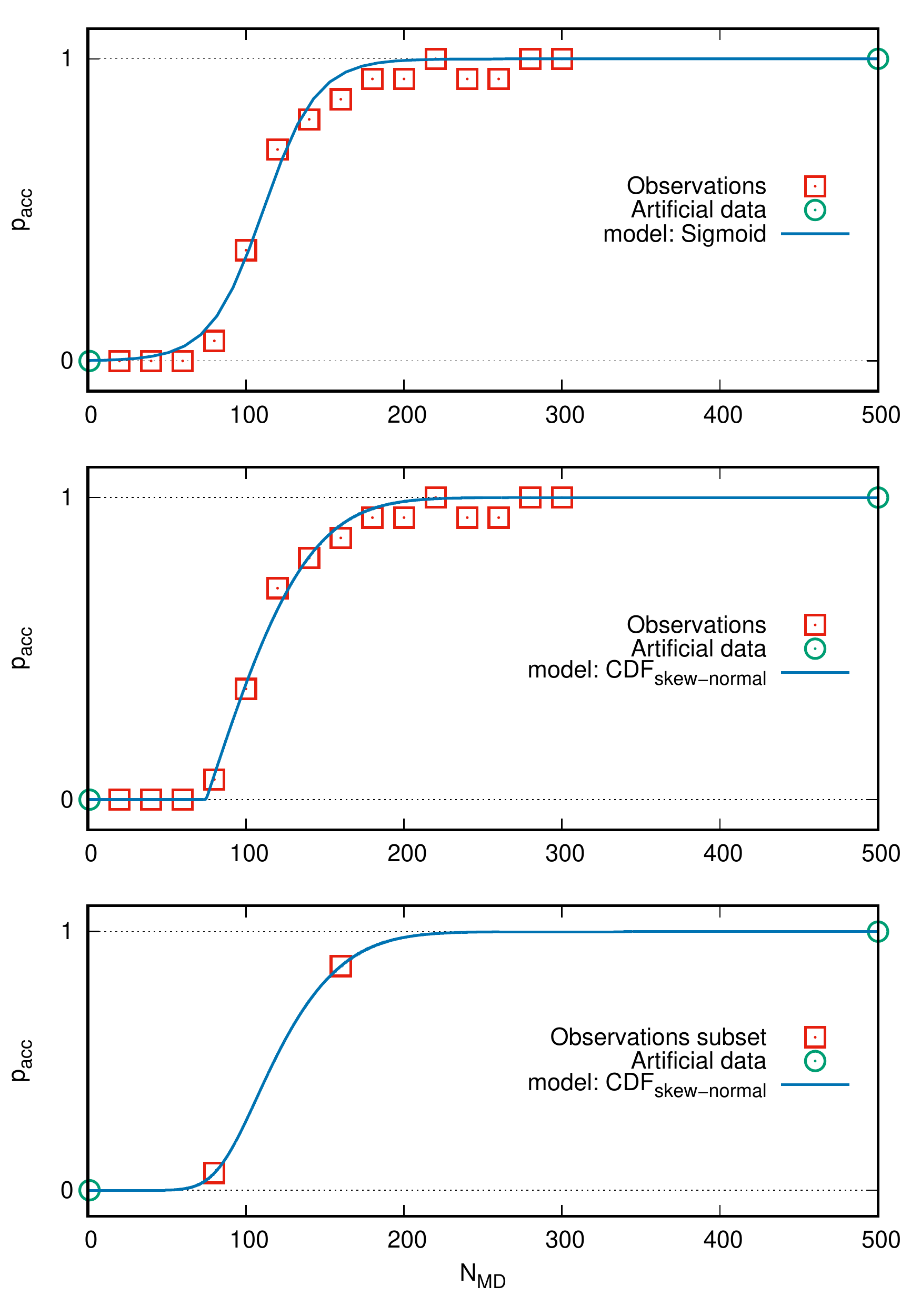}
	\caption{The model based on the CDF of the skew-normal distribution (middle) seems to fit the real observations better than a sigmoid function (top). By providing two extra artificial points, the model gives a reasonable shape even if a small subset of the measurements is available (bottom).}
	\label{cdfshape}
\end{figure}

\paragraph {Second phase -- creating a model}
The design of the second phase was based on visual observations from the shapes of the relationship of $p_\text{acc}$ versus $N_\text{MD}$. The resulting figures are reminiscent of the Sigmoid function, such as a logistic function, not only in shape, but also in the range of the dependent variable, which is between 0 and 1 in both cases. These are important pieces of information that we have provided to the model in order to make it more accurate and representative, even with a very small number of training data. For our purposes it seemed that the most appropriate solution was to fit the Cumulative Distribution Function (CDF) of the skew--normal distribution~\cite{azzalini1985class}, which is not always symmetric as a Sigmoid function. The skew--normal distribution is a generalization of the normal distribution that allows non--zero skewness. This impacts the shape of its CDF accordingly, giving our model the ability to change the amount of steepness in the two `bend points' of its `S'--like shape (see fig.~\ref{cdfshape} for an example for modelling real data).

The auto--tuner is written in python and uses the numpy and scipy libraries~\cite{walt2011numpy}, as well as the python version of the `sn' package~\cite{sn} for calculating the CDF of the skew--normal distribution for different parameters. The fitting of the model is done by calling the scipy.optimize.least\_squares function of scipy, that implements the Levenberg--Marquardt algorithm~\cite{more1978levenberg} for minimizing a loss function, `soft\_l1' in this case, which is a smooth approximation of the L\textsubscript{1} norm condition for robustness~\cite{triggs1999bundle} (see alg.~\ref{alg_nmd_tuner} for the pseudocode in~\ref{app:step_size}).\\

\subsection{Tuning the Hasenbusch parameters}
We now shift our aim to find a routine that optimizes the \HB\ parameters automatically. Using the auto--tuner of the previous section, we first thermalize the system with or without some fixed \HB\ parameters. After this we find appropriate staggered masses and time scales and, in the end, optimize $N_\text{MD}$ for this set of parameters using the HMC auto--tuner once more. 

\begin{figure}[h!]
	\centering
	\includegraphics[angle=270, width=0.9\textwidth]{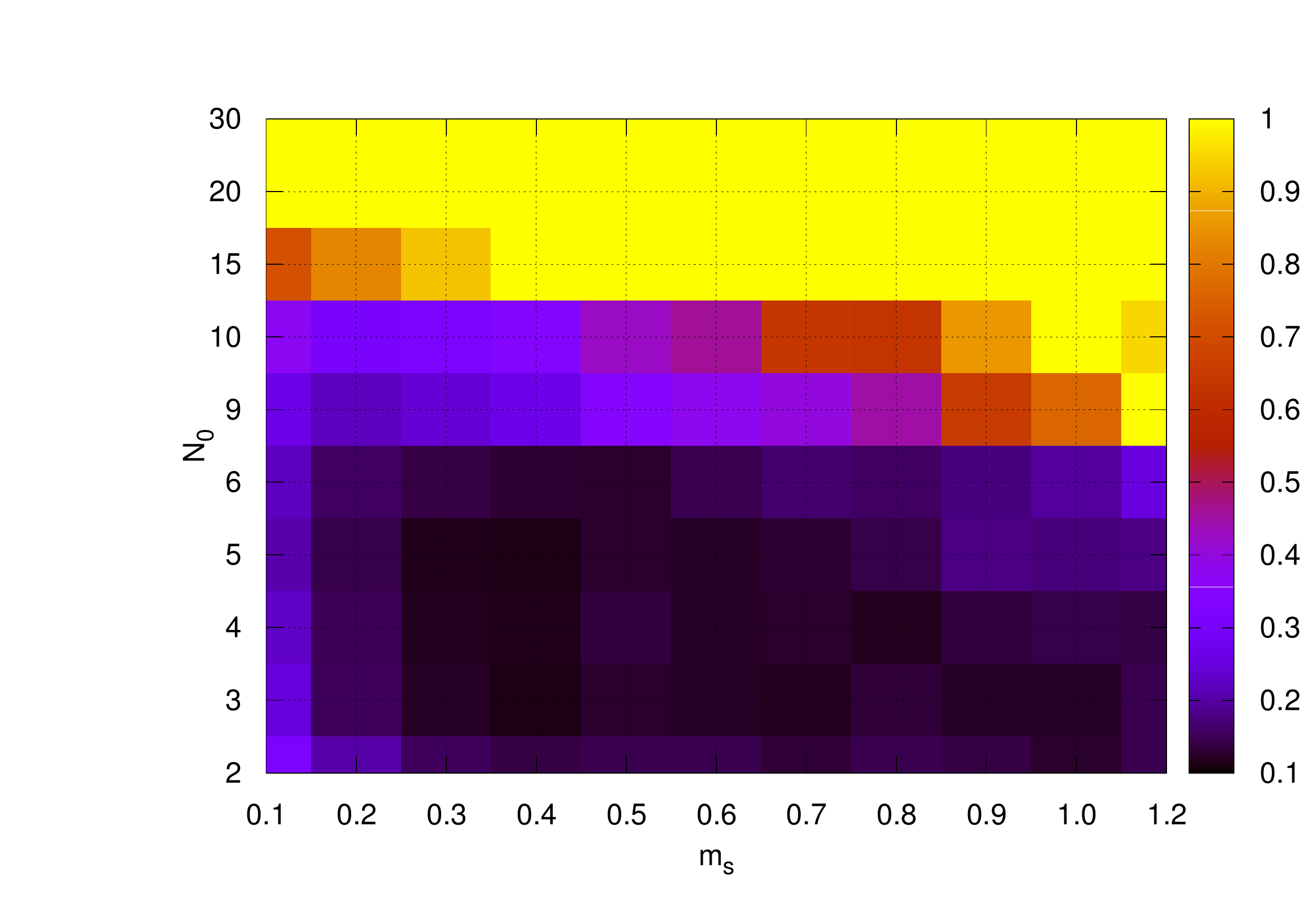}
	\caption{Runtime of one trajectory depending on the staggered mass $m_s$ and the ratio between the time scales $N_0$ normalized by the runtime without Hasenbusch acceleration and weighted by the square root of the error $\Delta\mathcal{H}$ made by the calculation. The weighting compensates that all calculations have been performed with the same $N_\text{MD}=540$. Values greater than 1 have been set to one 1 in this plot.}
	\label{time_weighted_256}
\end{figure}

Empirical data suggests a very broad minimum in parameter space (see fig.~\ref{time_weighted_256} for one staggered mass term). This means that the tuning does not have to be very accurate to achieve a speed that is nearly optimal. This is why we do not attempt to find the exact minimum of the CPU--time per trajectory. Instead we provide a simple procedure that can be automatized to find a set of parameters good enough for any practical purpose.

For the systems considered until now we did not see any significant improvements in using more than two \HB\ terms. Because of this presently our auto--tuner uses exactly two staggered masses, but it can be easily generalized to more masses should the need arise. Since the number of mass terms is two, the masses themselves have to be calculated. It seems reasonable to distribute them equidistantly in the spectrum, where equidistantly has to be understood in a geometrical sense. For $n$ masses this leads to
\begin{equation}
\mu_i=\lambda_\text{min}\left(\frac{\lambda_\text{max}}{\lambda_\text{min}}\right)^{\frac{i+1}{n+1}}\label{mu_i_eqn}
\end{equation}
with $\lambda_\text{min,max}$ gained from a fit of the solver time. The time scales then can be found knowing that we use a second order integrator and thus the error of the Hamiltonian $\Delta \mathcal{H}$ made in one trajectory scales approximately with
\begin{equation}
\Delta \mathcal{H}\propto\sum_i \left|F_i\right|N_i^2
\end{equation}
where $i$ counts all forces introduced above. Because of symmetry this term is minimized near the point where all addends have the same value. Thus we get a first approximation for the time scale ratios via
\begin{equation}
N_i'=\sqrt{\frac{F_{\phi+\chi_n}\!\left(\mu_{n-1}\right)}{F_{\chi_i}\!\left(\mu_{i-1},\mu_i\right)}}\,.\label{approx_N_i_eqn}
\end{equation}
For two masses one finds that $N_0'\approx N_1'$ independently of the system as long as $c\approx 0.4$. As $N_1$ has to divide $N_0$ the only appropriate choice is $N_0=N_1$. One finds that in most cases the results are best if one chooses
\begin{equation}
N_0=N_1=\left\lfloor\sqrt{N_0'\cdot N_1'}\right\rfloor\label{used_N_eqn}
\end{equation}
instead of rounding.\\
In summary the combined auto--tuner works as follows:
\begin{enumerate}
	\item Thermalize the system with some fixed \HB\ parameters (one staggered mass with $m_s=0.5$, $N_0=1$ works quite good) and find a good value for $N_\text{MD}$.
	\item Calculate solver time and forces for some different mass terms.
	\\We calculate with $N_0=N_1=1$ for the tuples\\$(m_0,\,m_1)\in \left\{(0.05,\,0.2),\,(0.1,\,0.3),\,(0.2,\,1.5),\,(0.4,\,0.8),\,(1.2,\,2)\right\}$.
	\item Fit the results according to the eqs.~\ref{solver_time_eqn},~\ref{small_force_eqn}, and~\ref{large_force_eqn}.
	\item Calculate the two \HB\ masses according to eq.~\ref{mu_i_eqn} and the time scales according to~\ref{approx_N_i_eqn} and~\ref{used_N_eqn}.
	\item Find the new optimum for $N_\text{MD}$.
\end{enumerate}

\section{Results\label{sec_results}}
The data presented here was produced at $\beta\kappa=8$ and $U/\kappa=2.5$ on single nodes $2\,\times\,$Intel(R) Xeon(R) CPU E5-2680, $\SI{40}{MB}$ Cache per node. The exact numbers deviate for other choices of parameters\footnote{Smaller values of $U/\kappa$ lead to significantly faster calculations and reduce the advantage of the presented algorithms compared to a standard CG solver. Yet even at $U/\kappa=0.5$ our optimized algorithm reaches a speedup of factor three to four. For $U/\kappa >2.5$ we find only small deviations from the presented data. Explicitly $U/\kappa=5$ has runtime and speedup deviations of less than $\SI{25}{\%}$ from the $U/\kappa=2.5$ case.} and on other computers but the qualitative behaviour is representative.

\begin{figure}[h!]
	\centering
    \includegraphics[angle=270, width=0.8\textwidth]{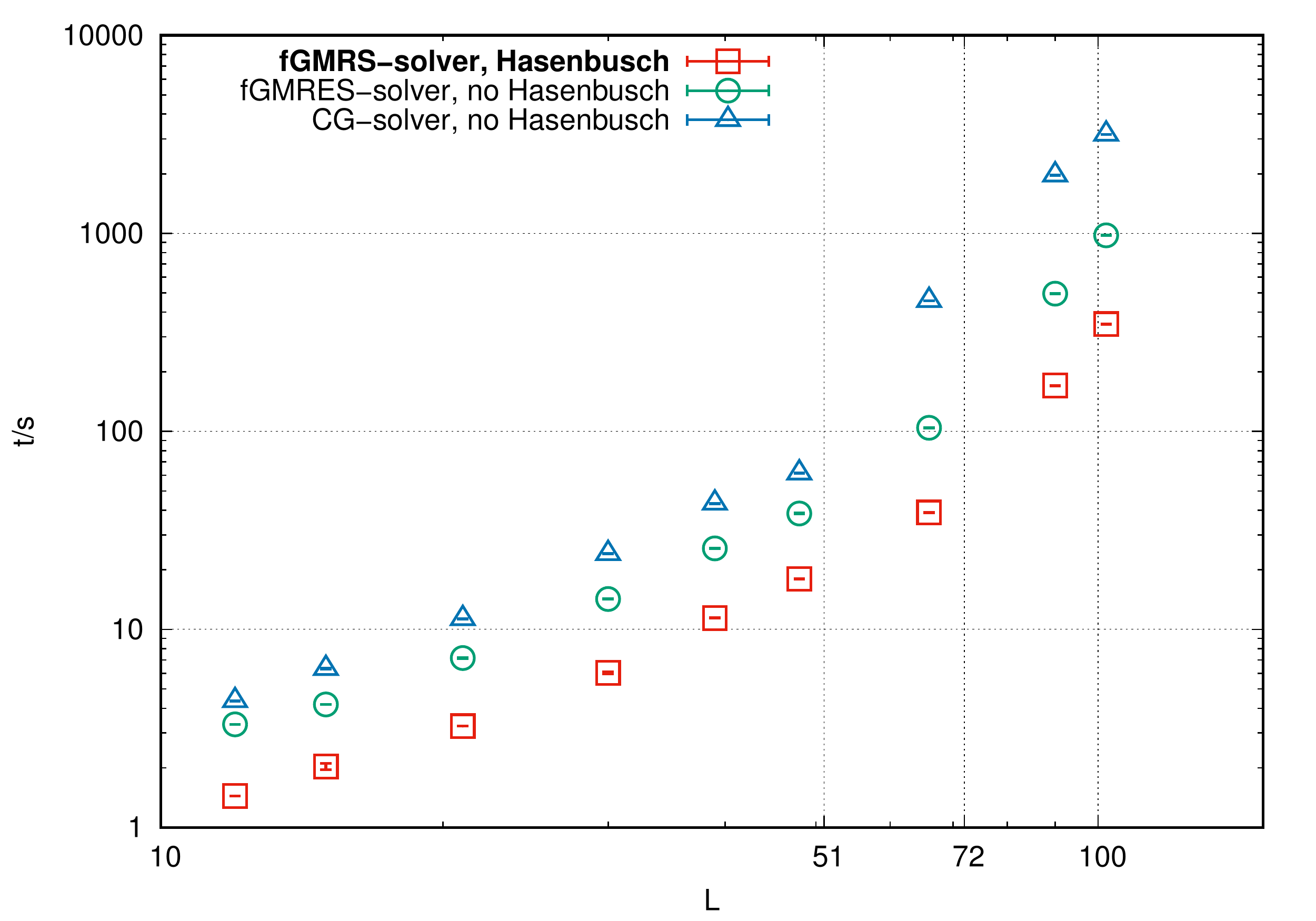}
	\includegraphics[angle=270, width=0.8\textwidth]{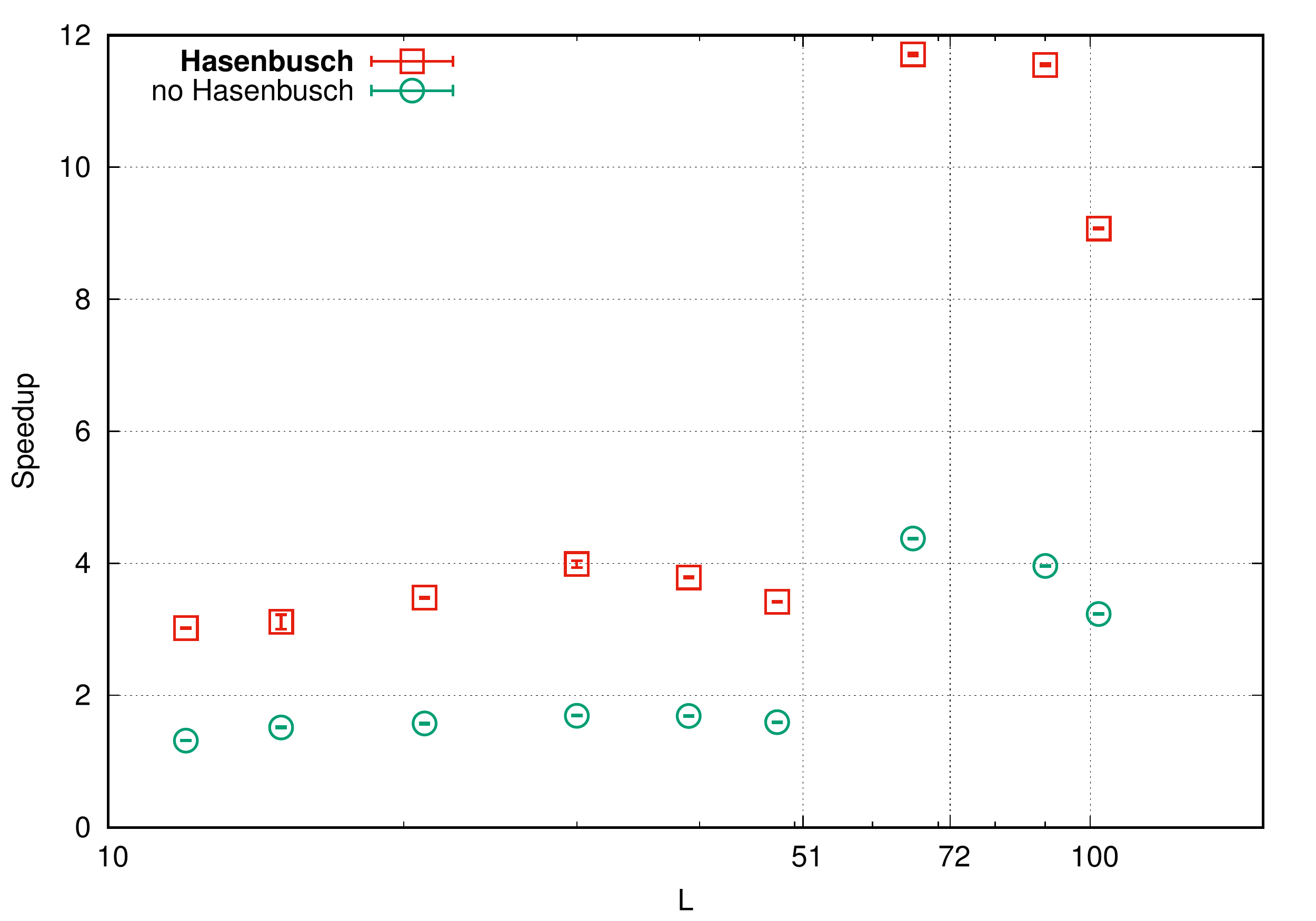}
	\caption{Top: Time needed per trajectory by the different algorithms for graphene sheets with $(L\times L)$ unit cells and $N_t=64$. Bottom: Speedup over the standard HMC algorithm (using double precision CG) for graphene sheets with $(L\times L)$ unit cells and $N_t=64$. The double precision CG solver exceeds L3 capacity at $L=51$, whereas the spCG exceeds L3 capacity at $L=72$. Our best algorithm is highlighted in bold.}
	\label{fig_time_l_dep_graphene}
\end{figure}

\begin{figure}[h!]
	\centering
  	\includegraphics[angle=270, width=0.9\textwidth]{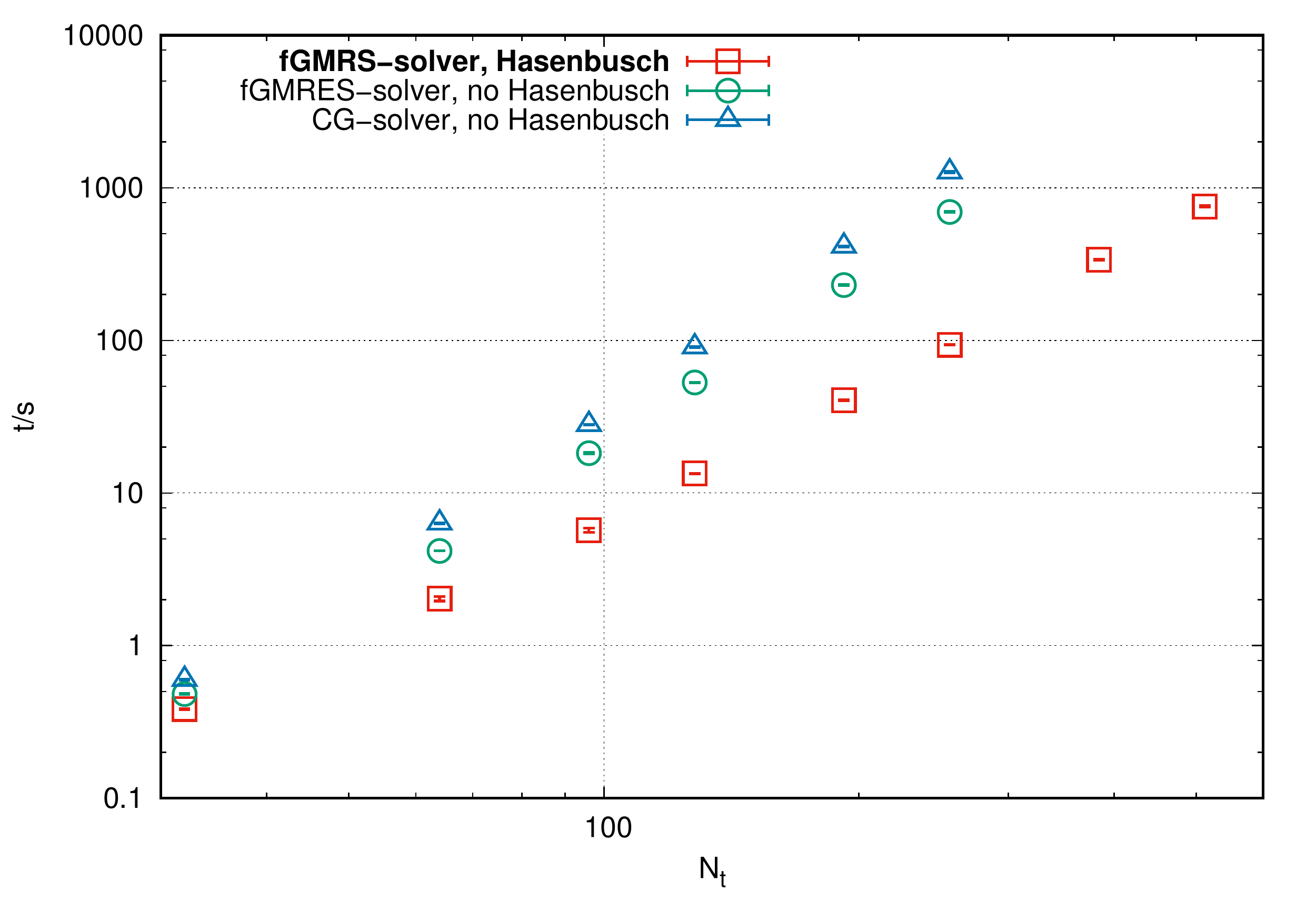}
	\includegraphics[angle=270, width=0.9\textwidth]{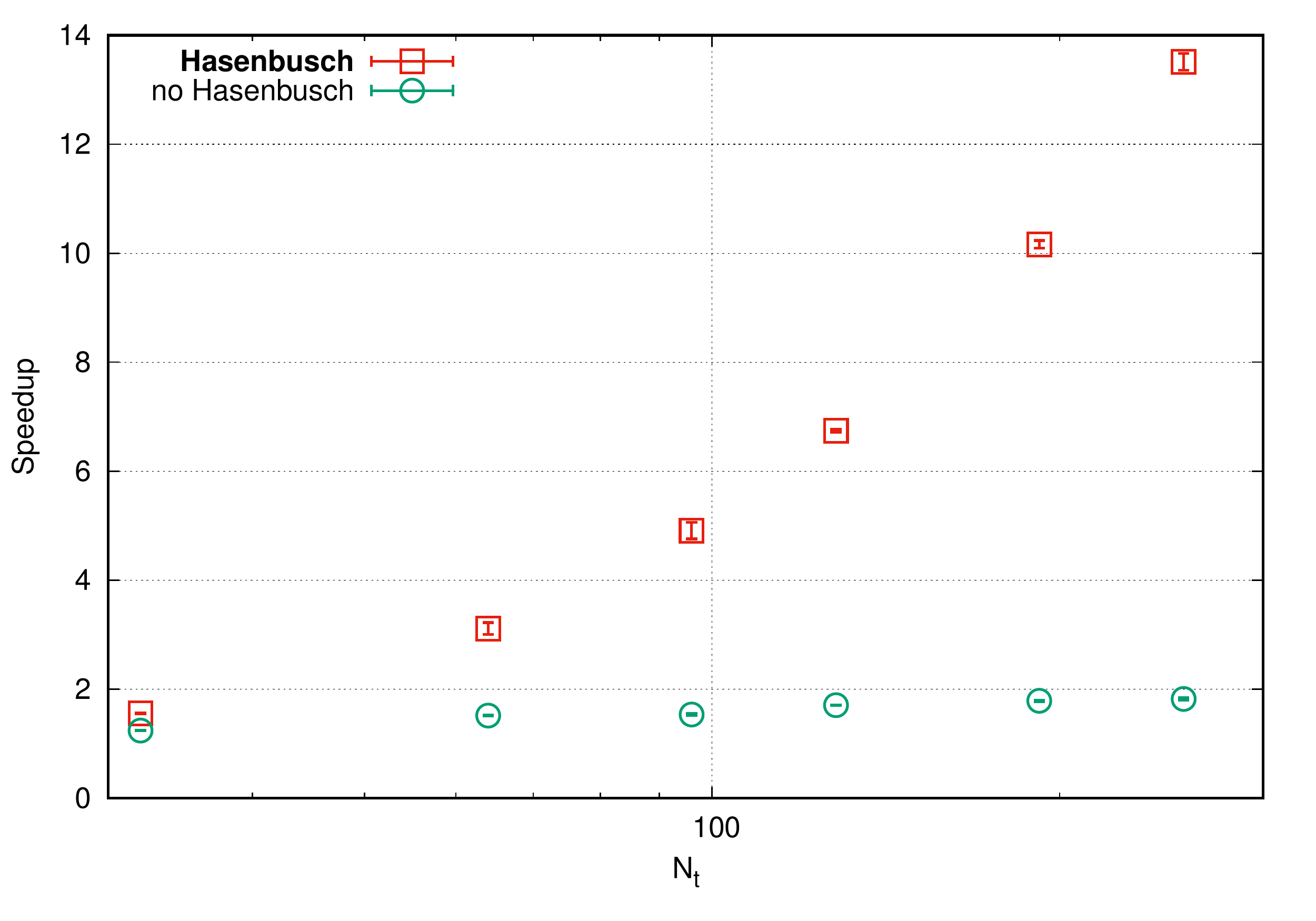}
	\caption{Top: Time needed per trajectory by the different algorithms for graphene sheets with $(15\times 15)$ unit cells depending on $N_t$. Bottom: Speedup over the standard HMC algorithm (using double precision CG) for graphene sheets with $(15\times 15)$ unit cells depending on $N_t$. Our best algorithm is highlighted in bold.}
	\label{fig_time_nt_dep_graphene}
\end{figure}

The auto--tuner as explained above has been implemented and was stable for all systems of \cnt s as well as graphene tested up to this point. All the results (with and without \HB\ acceleration) presented in this section have been tuned in this way. Figures~\ref{fig_time_l_dep_graphene} and ~\ref{fig_time_nt_dep_graphene} visualize the speedup depending on the system size, achieved for one trajectory by the introduction of the fGMRES solver and additional \HB\ acceleration respectively. \added{The lengths to which the trajectories have been tuned can be found in table~\ref{tab_nmds}.}

\begin{table}[h]
\centering
\begin{tabular}{r|r|r|r|r|r|S[round-mode=places,round-precision=2]|S[round-mode=places,round-precision=2]}
	&& CG-solver & fGMRES-solver & \multicolumn{4}{c}{fGMRES-solver, Hasenbusch}\\
    $L$ & $N_t$ & $N_\text{MD}$ & $N_\text{MD}$ & $N_\text{MD}$ & $N_\text{MD}/N_0$ & \multicolumn{1}{c}{$\mu_0$} & \multicolumn{1}{c}{$\mu_1$}\\\hline
        12 & 64 & 34 & 35 & 18 & 9 & 0.514011 & 1.61581 \\
        15 & 32 & 12 & 12 & 8 & 4 & 0.416651 & 0.729772 \\ 
        15 & 64 & 37 & 35 & 18 & 9 & 0.389511 & 1.18414 \\ 
        15 & 96 & 72 & 73 & 26 & 13 & 0.330014 & 1.36657 \\ 
        15 & 128 & 126 & 124 & 44 & 22 & 0.510381 & 2.36384 \\
        15 & 192 & 269 & 263 & 70 & 35 & 0.497538 & 2.27238 \\
        15 & 256 & 464 & 474 & 98 & 49 & 0.413291 & 2.09108 \\
        21 & 64 & 38 & 39 & 22 & 11 & 0.542087 & 1.84349 \\ 
        30 & 64 & 41 & 43 & 20 & 10 & 0.414843 & 1.11987 \\
        39 & 64 & 48 & 49 & 30 & 10 & 0.325614 & 0.738415 \\
        48 & 64 & 45 & 47 & 30 & 10 & 0.328139 & 0.704837 \\
        66 & 64 & 54 & 61 & 24 & 12 & 0.41166 & 0.940914 \\ 
        90 & 64 & 64 & 68 & 26 & 13 & 0.388389 & 0.985207 \\
        102 & 64 & 66 & 63 & 30 & 15 & 0.4054 & 1.09624
\end{tabular}
\caption{\added{Lengths of the auto--tuned trajectories for the different algorithms. In the case of the Hasenbusch-algorithm in addition the two staggered masses and their time scale are given. The third column from the right gives the length of the trajectory on the most coarse time scale.}}\label{tab_nmds}
\end{table}

We analysed the runtime scaling with the system size. One finds a monomial dependence of the time on $L$ and $N_t$ in some regions but not globally. The dependence in the spatial size $L$ (see fig.~\ref{fig_time_l_dep_graphene}) is comparable for all algorithms but the algorithm including \HB\ acceleration and the new fGMRES solver is much faster than the algorithm without \HB\ acceleration but with mixed precision solves, which in turn is significantly faster than the algorithm without \HB\ acceleration using only the double precision CG solver. In addition one encounters a jump in the runtime between $L=48$ and $L=66$ for the original algorithm and between $L=66$ and $L=90$ for the algorithms using the spCG preconditioner. At these points the memory needed by the solver exceeds the L3-cache of $\SI{40}{MB}$. After this threshold the scaling of the runtime becomes stronger and the ratio between the single precision and the double precision algorithms increases dramatically. It is expected that this advantage is going to decrease as the system size grows even larger because L3-cache is not going to play an important role any more neither for double nor single precision solver. We see this behaviour already at $L=102$, so we are not going to simulate yet larger systems on single node in recent future.

The L3-threshold is not reached in figure~\ref{fig_time_nt_dep_graphene}, but here one can see that the scaling of the runtime with $N_t$ is much stronger without \HB\ acceleration. Not only the runtime but also the tuning takes much longer without \HB\ acceleration regardless of the much lower amount of trajectories that have to be calculated. This is why we can present results for systems as large as $N_t=512$ with \HB\ acceleration but only $N_t=256$ without. Our experience up to this point leads us to conclude that the continuum limit ($N_t\rightarrow\infty$) is not only more expensive but also much more crucial for physical results than the infinite volume limit ($L\rightarrow\infty$), making this advantage of the \HB\ algorithm rather important.

\subsection{Benchmarking to the 4-site problem\label{sect:benchmarking}}
\begin{figure}
\centering
\includegraphics[angle=270,width=.8\columnwidth]{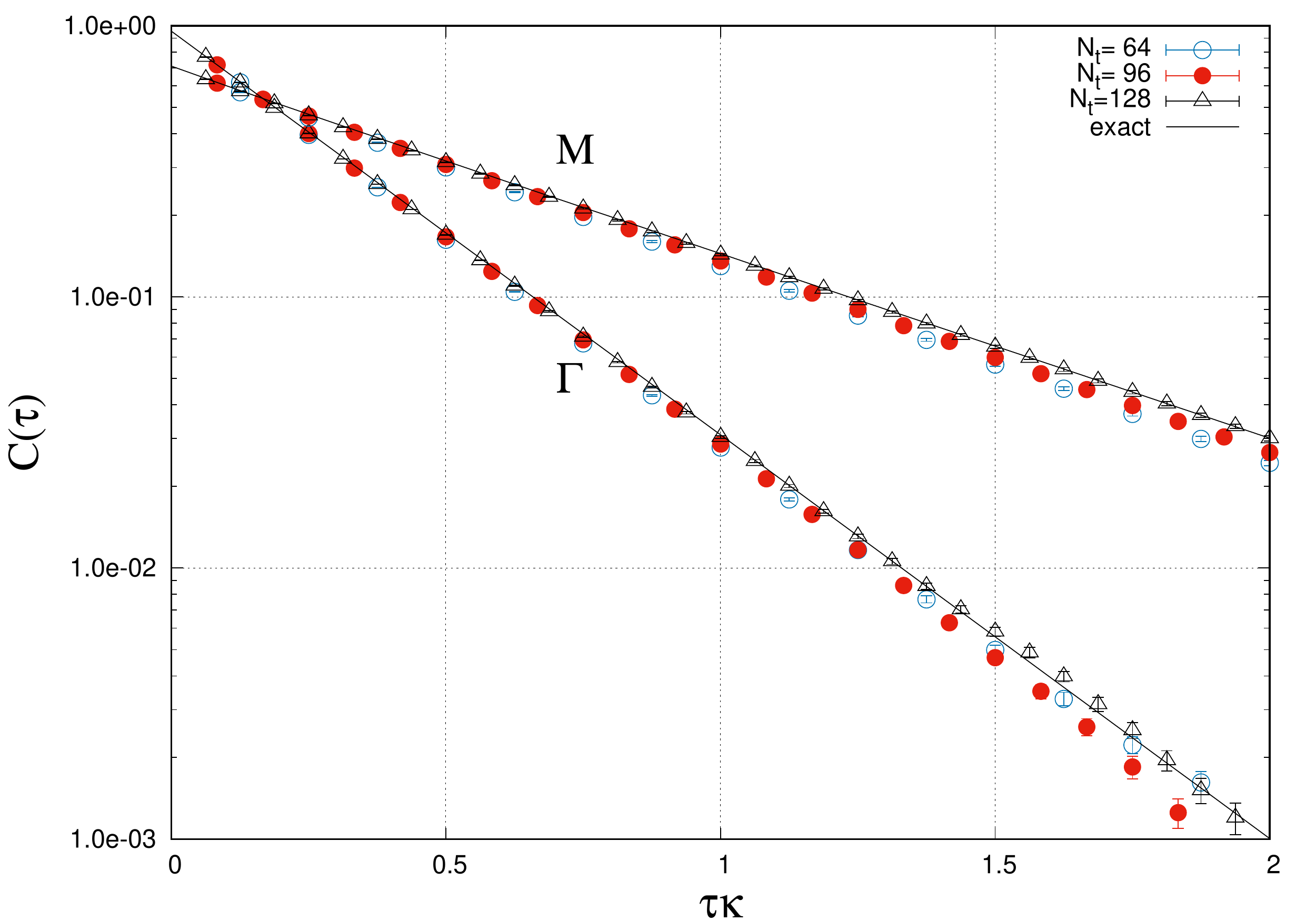}
\caption{The two correlators (labelled M and $\Gamma$) calculated from a 4-site model with $U/\kappa=4$ and $\beta\kappa=8$. The number of timesteps $N_t$ is given in the figure.  Uncertainties are too small to be seen in many cases.  The black lines are the exact results.  \label{fig:benchmark results}}
\end{figure}

To ensure that our implementation of the auto--tuner and the \HB\ mass shifts produce correct physics, we compute the allowed correlators for the 4-site model with nearest neighbours drawn from a cubic geometry (i.e. hexagonal lattice), and compare to the corresponding correlators obtained from direct diagonalization of the Hubbard Hamiltonian.  We simulate with a Hubbard ratio $U/\kappa=4$,  $\beta\kappa=8$, and $N_t=64$, 96, 128, and 160.  Figure~\ref{fig:benchmark results} shows our calculations, excluding the $N_t=160$ results to reduce clutter.  Here the expected convergence of the correlators as we increase $N_t$ is observed.  

We can extract the interacting quasi-particle eigenenergies of this system by fitting an exponential to these correlators, as the expected behaviour of these correlators is $C(\tau)\sim e^{-E\tau}$ for large $\tau$.  In practice, we fitted exponentials within a given time window.  In particular, $E_\Gamma$ was fitted to an exponential in the time window $\tau\kappa\in[.2,1]$, whereas $E_M$ was fitted in $\tau\kappa\in[.6,2]$.  In \autoref{tab:fit results} we show these extracted energies as well as the exact results from direct diagonalization. As these results were only for benchmarking, we did not optimize our fit windows when performing our fits.  Still, we find excellent agreement with the exact results for $N_t=128$ and higher. 
\begin{table}
\centering
\caption{Extracted energies from correlators of the 4-site model, as a function of $N_t$.  $E_\Gamma$ was extracted by fitting an exponential to the correlator in the time window $\tau\kappa\in[.2,1]$.  $E_M$ was extracted from exponential fits in window $\tau\kappa\in[.6,2]$. Uncertainties were obtained from a bootstrap ensemble of fits. The last column gives the exact results for direct diagonalization of the Hubbard Hamiltonian.\label{tab:fit results}}
\begin{tabular}{c|c|c|c|c|c}
$N_t$ & 64 & 96 & 128 & $160$ & exact\\
\hline
$E_\Gamma/\kappa$ & \num{3.555\pm 0.016} & \num{3.501\pm 0.011} & \num{3.445\pm 0.012} & \num{3.447\pm 0.014} & 3.429\\
$E_M/\kappa$ & \num{1.647\pm 0.023} & \num{1.629\pm 0.021} & \num{1.542\pm 0.018} & \num{1.567\pm 0.022} & 1.565
\end{tabular}
\end{table}

\section{Conclusion\label{sect:conclusions}}
In this paper we investigated different attempts at accelerating lattice Monte Carlo simulations of the Hubbard model on a hexagonal lattice.  We implemented higher order symplectic integrators in our MD integrations, and introduced mixed-precision CG solvers as a `preconditioner'.  In the former case we found the Omelyan integrator to be competitive with the force gradient method, while in the latter the mixed precision preconditioner gave a speedup of approximately 1.9.  In large systems, such as $66\times 66$ or larger, the introduction of the fGMRES solver also provided a factor of $\sim$2.2 speedup.

Our greatest speedup came with the implementation of \HB\ acceleration via mass shifts.  With just two \HB\ masses, and in combination with the fGMRES solver, we were able to get a speedup of well over an order of magnitude for lattices of size $90\times 90$ and higher.  

The numerous parameters available in our simulations required the use of an auto--tuner to find their optimal values.  By establishing scaling relations between the \HB\  masses and forces, as well as between the acceptance rate and number of MD steps, we devised an efficient and stable procedure for optimizing the parameter sets in our lattice MC simulations to a target 66\% acceptance rate. 

To ensure that our implementation of the \HB\  acceleration and auto--tuner performed correctly, we simulated the 4-site system and compared our results to direct diagonalization, finding excellent agreement.  

Our increased speedup now allows us to simulate hexagonal lattices of unprecedented sizes.  With only single-node resources, we were able to simulate up to $102\times 102$ lattices with $N_t=64$.  Results with even larger spatial sizes and $N_t$ are in preparation. If optimized over multi-nodes, we anticipate the ability to simulate on systems an order of magnitude in size larger, thereby approaching system sizes comparable to physical applications.  We are currently developing this capability.    

\section*{Acknowledgements}
We thank E. Berkowitz, C. K\"orber, T. L\"ahde, and J.-L. Wynen for their insightful discussions related to this work. We also thank M. Ulybyshev for his comments.  This work is supported in part by the DFG grants SFB TRR-55 and SFB TRR-110 (Sino German CRC). The simulations were carried out using computing resources which received funding from the European Community's Seventh Framework Programme (FP7/2007-2013) under Grant Agreement n$^\circ$~287530 (DEEP) and 610476 (\mbox{DEEP-ER}), and from the Horizon~2020 Programme (H2020-FETHPC) under Grant Agreement n$^\circ$~754304 (\mbox{DEEP-EST}). P. Papaphilippou was supported by the Summer of HPC programme of the PRACE-5IP project (2017-2019, RI-730913) carried under EC Research and Innovation Action of  the H2020 Programme. The present publication reflects only the authors' views. The European Commission is not liable for
any use that might be made of the information contained therein.

\appendix

\section{Implementation of the Flexible Generalized Minimal Residual solver}\label{app_fgmres}
The algorithm of the flexible GMRES from~\cite{fgmres} with the spCG as preconditioner in our implementation can be found in algorithm~\ref{alg_fgmres}
\begin{algorithm}
	\SetKwInOut{Input}{input}\SetKwInOut{Output}{output}
	\Input{\texttt{TOL}, $b$, $A$}
	\Output{$x$}
	\BlankLine
	solve $A\cdot x = b$ for $x$ using spCG\;
	$r = b-A\cdot x$\;
	$\text{\texttt{ERR}} = |r|$\;
	$w = r$\;
	\While{$\text{\texttt{ERR}} > \text{\texttt{TOL}}\cdot |b|$}{
		\For{$j=0,\dots,\,m-1$}{
			\If {$\text{\texttt{ERR}} < \text{\texttt{TOL}}\cdot |b|$}{
				break\;
			}
			$v_j = w/|w|$\;
			\If {$\text {\texttt{TOL}}\cdot a\cdot |b|/\text {\texttt{ERR}} < 1$}{
				solve $A\cdot z_j = v_j$ for $z_j$ using spCG with $\text {\texttt{TOL}} = \text {\texttt{TOL}}\cdot a\cdot |b|/\text {\texttt{ERR}}$\;
			}\Else{
				$z_j = v_j$\;
			}
			$w = A\cdot z_j$\;
			$h_{ij} = w\cdot v_i$\;
			$w = w - \mathrm{sum}\left(h_{ij}\cdot v_i, i = 0,\dots,\,j\right)$\;
			$h_{j+1,j} = |w|$\;
			$\text {\texttt{ERR}} = \text {\texttt{ERR}}\cdot |w|$\;			
		}
		$y = \mathrm{argmin}\left|\left(|r|,0,\dots,0\right) - h\cdot y\right|$\;\label{argmin}
		$x = x + \mathrm{sum}\left(y_i\cdot z_i,\: i = 0,\dots,\,j\right)$\;
		$r = b-A\cdot x$\;
		$\text {\texttt{ERR}} = |r|$\;
		$w = r$\;
	}
	\SetAlgoRefName{A.1}
	\caption{Flexible GMRES}\label{alg_fgmres}
\end{algorithm}
with some relative tolerance \texttt{TOL}, the complex vector $b\in\mathbb{C}^n$ and the matrix $A\in\mathbb{C}^{n\times n}$ as input parameters and $x,r,w\in\mathbb{C}^n$, $v,z\in\mathbb{C}^{m\times n}$, $h\in\mathbb{R}^{m+1\times m}$, $y\in\mathbb{R}^m$. Here $|\cdot|$ denotes the euclidean norm. The minimization in line~\ref{argmin} has been performed by solving $h^Th\,y=|r|\cdot h^T\,e_1$ for $y$ using a Cholesky decomposition. $m=10$ has been chosen as the restart parameter and it showed to be good to set $a=5$ in the tolerance of the spCG preconditioner. 

\section{Tuning the HMC step size}\label{app:step_size}
In alg.~\ref{alg_nmd_tuner} we can see the pseudocode for the auto--tuner of the HMC step size. It can be seen as a single routine that takes a parameter file for all parameters of the simulator, ignores the provided \replaced{$\text{N\textsubscript{MD}}=\frac{1}{\Delta t}$}{N\textsubscript{MD}}, runs a number of simulations with automatically set N\textsubscript{MD} values and stops when the final N\textsubscript{MD} value is expected to continue yielding a mean trajectory acceptance rate of 0.66. \\

\begin{algorithm}[H]
\LinesNumbered
  \KwResult{The N\textsubscript{MD} for which the acceptance rate is close to 0.66}
 N\textsubscript{MD} $\leftarrow$ 500\;
 initialize or resume from a previous experiment\;
\tcc{Main loop}
 \While{True}{
\tcc{1st phase - minimizing the confidence interval}
  trajectories $\leftarrow$ 0\;
  \While{conf. interval $>$ 0.25 {\bf or} trajectories $\leq$ 5}{
    p\textsubscript{acc} $\leftarrow$ simulate(N\textsubscript{MD})\;
    trajectories $\leftarrow$ trajectories + 1\;
  }
  Append N\textsubscript{MD} to N\textsubscript{MD}list\;
  Append mean p\textsubscript{acc} to p\textsubscript{acc}list\;
\tcc{2nd phase - building a model}
{
   training\_set $\leftarrow$ (N\textsubscript{MD}list, p\textsubscript{acc}list)\;
   fit\_function $\leftarrow$ \textit{f(x) = CDF\textsubscript{skew--normal}($\beta$\textsubscript{1}x + $\beta$\textsubscript{0}, $\alpha$) + $\epsilon$} \; 
   model\_build(training\_set, fit\_function)\;
   N\textsubscript{MD} $\leftarrow$ model\_predict\_N\textsubscript{MD}\_from\_p\textsubscript{acc}(0.66)\;
   \tcc{Stopping conditions - convergence heuristic}
     \If{count total trajectories $\geq$ 500 {\bf or } N\textsubscript{MD}list.count(N\textsubscript{MD}) $\geq$ 3 {\bf or } times 0.66 was in conf. interval $\geq$ 3 }{
   	 break\;
  }
  }
 }
 \SetAlgoRefName{B.1}
\caption{High-level description of the HMC step size auto--tuner}\label{alg_nmd_tuner}
\end{algorithm}


\bibliography{mybibfile}

\end{document}